\documentclass[
superscriptaddress,
article
amsmath,amssymb,
aps,
]{revtex4}

\usepackage[utf8]{inputenc}
\usepackage{amsmath}
\usepackage[T1]{fontenc}
\usepackage{amsmath,amsfonts,amssymb,amsthm,epsfig,epstopdf,url,array}
\usepackage{bm}
\usepackage{graphicx}
\usepackage{mathtools}
\usepackage{physics}
\usepackage{comment}
\usepackage{color}

\usepackage{dcolumn}
\usepackage{bm}
\usepackage{graphicx}
\usepackage{comment}

\newcommand{\bea}{\begin{eqnarray}}

\newcommand{\eea}{\end{eqnarray}}

\usepackage{color}

\newcommand{\blue}{\textcolor{black}}

\let\oldref\ref
\renewcommand{\ref}[1]{(\oldref{#1})}

\begin{document}

\title{\blue{Exact extreme, order and sum statistics in a class of strongly correlated system}}
\author{Marco Biroli}
\affiliation{LPTMS, CNRS, Univ.  Paris-Sud,  Universit\'e Paris-Saclay,  91405 Orsay,  France}

\author{Hernán Larralde}
\address{Instituto de Ciencias Físicas, UNAM, Av. Universidad s/n, CP 62210 Cuernavaca Morelos, México}

\author{Satya N. Majumdar}
\affiliation{LPTMS, CNRS, Univ.  Paris-Sud,  Universit\'e Paris-Saclay,  91405 Orsay,  France}

\author{Gr\'egory Schehr}
\affiliation{Sorbonne Universit\'e, Laboratoire de Physique Th\'eorique et Hautes Energies, CNRS UMR 7589, 4 Place Jussieu, 75252 Paris Cedex 05, France}

\date{\today}

\begin{abstract}
\blue{Even though strongly correlated systems are abundant, only a few exceptional cases admit analytical solutions. In this paper we present a large class of solvable systems with strong correlations.}. We consider a set of $N$ independent and identically distributed (i.i.d) random variables
$\{X_1,\, X_2,\ldots, X_N\}$ whose common distribution has a parameter $Y$ (or a set of
parameters) which itself is random with its own distribution. For a fixed value of this
parameter $Y$, the $X_i$ variables are independent and we call them conditionally
independent and identically distributed (c.i.i.d). However, once integrated over the
distribution of the parameter $Y$, the $X_i$ variables get strongly correlated,
yet retaining a solvable structure for various observables, such as for the sum and
the extremes of $X_i$'s. This provides a simple \blue{procedure} to generate a class of
solvable strongly correlated systems. We illustrate how this \blue{procedure} works via
three physical examples where $N$ particles on a line perform independent (i) Brownian
motions, (ii) ballistic motions with random initial velocities, and (iii) L\'evy flights, but
they get strongly correlated via {\it simultaneous resetting} to the origin. Our results are
verified in numerical simulations. This \blue{procedure} can be used to generate an endless variety of
solvable strongly correlated systems.
\end{abstract}

\date{May 2022}

\maketitle




\section{Introduction}

The study of the statistics of functions of independent random variables is central to the foundations of probability theory. In fact, the central limit theorem, which describes the sum of independent random variables, long predates modern probability theory and has led to innumerable applications and results \cite{Fischer}. Another observable of much interest is the statistics of the maximum (or the minimum) of a set of random variables. This is usually referred to as ``extreme value statistics'' (EVS). The EVS of independent random variables are also well understood \cite{Gumbel, Leadbetter, Fortin, Arnold, Nagaraja} and are of a particular practical interest since, although rare, being extreme by nature these events may lead to devastating consequences. Indeed, in many contexts, such as engineering \cite{Gumbel}, environmental sciences \cite{Katz}, computer science \cite{Krapivski_00, Majumdar_00, Majumdar_02, Majumdar_03, Majumdar_05}, finance \cite{Majumdar_08, Embrecht} or physics \cite{Fortin, Derrida_81, Bouchaud_97}, to cite but a few, the understanding of extreme events is a matter of crucial importance. For example, in engineering and environmental sciences the statistics of rare events, such as weak components or natural disasters, may be of major importance to the endeavor at hand. Interestingly, many of these statistics tend to universal asymptotic forms in the limit of large number of variables. Examples are, of course, the Central limit theorem and the L\'evy limit distributions for the sums of independent random variables. In parallel, one of the most important contribution to the field of extreme-value theory was the Fisher-Tippett-Gnedenko theorem, which universally characterized the EVS of a set of identically distributed independent random variables. The distribution of the maximum in the large $N$ limit converges to the one of the three possible limiting forms: Gumbel, Fr\'echet or Weibull according to the behavior of the tail of the distribution of the variables. These results have been generalized for non identically distributed independent variables \cite{Weissman, Anderson_78, Anderson_84, Davis, Smith} and for a large set of weakly correlated identically distributed random variables, where the results, in the large $N$ limit, reduce to the independent case \cite{Majumdar_20}. However, only rare and punctual results are known for the statistics of strongly correlated random variables \cite{Majumdar_20, Derrida_85, Derrida_86, Tracy_94, Tracy_96, Dean_01,SatyaPaul03,MC_04,MC_05,GM_06,Bertin_06}. Yet, many physical systems present strong correlations and a lack of universal characterization for such systems greatly hinders progress in this area.\\

In a recent paper \cite{Biroli_23} we introduced a model of $N$ non-interacting Brownian motions on the line, that are subjected to simultaneous resetting to the origin. This simultaneous resetting makes the $N$ particles strongly correlated, yet it retains an exactly solvable structure. Indeed, many observables including the EVS have been computed explicitly \blue{for this model \cite{Biroli_23}}.  This \blue{straightforward} mechanism of generating strong correlations via simultaneous resetting can actually be generalised to a wider class of systems, as we show in this paper. The EVS of strongly correlated variables are, in general, very hard to solve and there exist only few exactly solvable cases~\cite{Majumdar_20, Derrida_85, Derrida_86, Tracy_94, Tracy_96, Dean_01, Bertin_06, MC_04, MC_05, GM_06}. This generalization that we provide here opens up a wider class of solvable models with strong correlations.

The mechanism behind this generalization is \blue{straightforward}. Consider a set of $N$ independent and identically distributed (i.i.d.) random variables $X_1, \cdots, X_N$ with a common distribution which contains a set of parameters \blue{$Y_1, Y_2, \cdots, Y_M = \vec{Y}$} which themselves are random variables with their own distribution. An example of the case $N=1$ and $M=1$ is where $X$ refers to the energy of a gas and $Y$ refers for example to the temperature or the magnetic field. In this case, the statistics of $X$, averaged over the distribution of $Y$, is referred to as "superstatistics" that has been studied in various contexts~\cite{superstat1,superstat2,superstat_vivo}. For fixed values of these parameters, the $X_i$ variables are statistically independent with a joint distribution  
\begin{equation}
{\rm Prob.}[X_1, \cdots, X_N | \blue{\vec{Y}}] = \prod_{i = 1}^N {\rm Prob.}[X_i | \blue{\vec{Y}}] \;. \label{P_joint_IID}
\end{equation}
Hence we call these $X_i$ variables conditionally independent and identically distributed (c.i.i.d.) random variables. However, when one integrates over the $Y_i$ variables, the joint distribution of $X_1, X_2, \cdots, X_N$ is no longer factorizable 
\begin{equation}\label{def:general}
{\rm Prob.}[X_1, \cdots, X_N] = \int \left\{\prod_{i = 1}^N {\rm Prob.}[X_i |\blue{\vec{Y}} ]\right\} {\rm Prob.}[\blue{\vec{Y}} ] ~\dd \blue{\vec{Y}} \;.
\end{equation}
Thus the $X_i$'s get correlated since they share the same set of parameters $\blue{\vec{Y}}$. 
A physical example of such a system is provided by \blue{a simplified version of} models of ``diffusing diffusivity'' where $N$ Brownian motions $X_1(t), \cdots, X_N(t)$ share a common diffusion constant $Y_1 = D$ \cite{CS2014,superstat_metzler} \blue{which does not evolve in time but is randomly initialized}. In this specific case, Eq. \ref{def:general} would read
\begin{equation}\label{eq:gauss_example}
{\rm Prob.}[X_1(t), \cdots, X_N(t)] = \int  \left\{\prod_{i = 1}^N \frac{1}{\sqrt{4 \pi D t}} \exp[ - \frac{X_i(t)^2}{4 D t} ]\right\} {\rm Prob.}[D] \; \dd D \;.
\end{equation}
This example clearly demonstrates that the joint distribution of $X_i$'s does not factorize and hence the $X_i$ variables are correlated. Many other physical examples motivated by stochastic resetting will be discussed in this paper\blue{, but this family of systems describes a wide variety of physical problems. We provide here a short non-exhaustive list of such problems so that the reader may grasp the type of problems which can be described by this protocol. This protocol can be used to describe: 
\begin{itemize}
\item the statistics of $N$ independent stochastic processes measured after a random time $Y = T$. 
\item the statistics of $N$ experimental observations $X_1, \cdots, X_N$ which depend on some experimental parameters $Y_1, \cdots, Y_M$ that may have some non-negligible uncertainties $\sigma_1, \cdots, \sigma_M$. Then, supposing we know ${\rm Prob.}[X_i | \vec{Y}]$, we can model ${\rm Prob.}[\vec{Y}]$ by a $M$ Gaussians centered around their expected values and with variances $\vec{\sigma}$.
\item the statistics of $N$ independent particles $X_1, \cdots, X_N$ evolving in an energy landscape $E(Y_1, \cdots, Y_M)$ which depends on some parameters $Y_1, \cdots, Y_M$ (magnetization, temperature, etc.) which are randomly initialized according to some distribution at $t = 0$.
\end{itemize}
}

A natural question is whether it is possible to generalize the well known results of certain observables for i.i.d. random variables, such as the sum or the EVS, to the c.i.i.d. variables. For example, for $N$ i.i.d. variables, it is known that the rescaled sum (sometimes referred to as the ``sample mean'') $C = \frac{1}{N} \sum_{i=1}^N X_i$ converges to a Gaussian random variable in the large $N$ limit: this is the celebrated Central Limit Theorem (CLT) \cite{Feller}. How does the CLT get modified for c.i.i.d. variables? Similarly, as mentioned before, the EVS of $N$ i.i.d. random variables, appropriately centered and scaled, converges for large $N$, to one of the three limiting distributions Gumbel, Fr\'echet and Weibull. Are there similar limiting universal distributions for c.i.i.d. variables? After recalling these results for i.i.d. variables in Section \ref{section:iid}, we will consider their generalisations to c.i.i.d. variables in Section \ref{ciid}. 

\begin{figure}
\centering
\begin{minipage}[b]{0.33\textwidth}
\centering
\includegraphics[width = \textwidth]{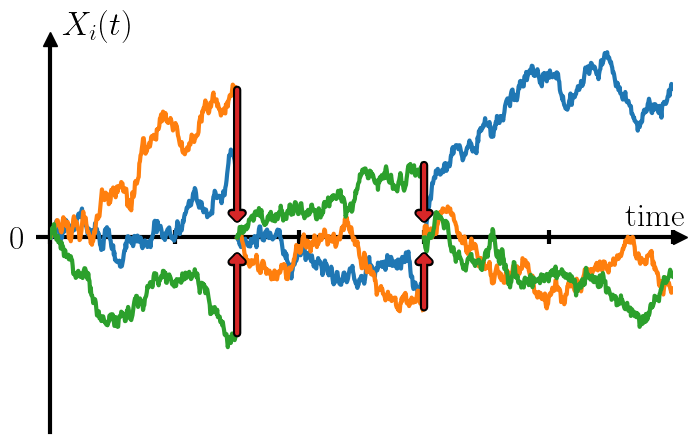}
\end{minipage}\hfill
\begin{minipage}[b]{0.33\textwidth}
\includegraphics[width=\textwidth]{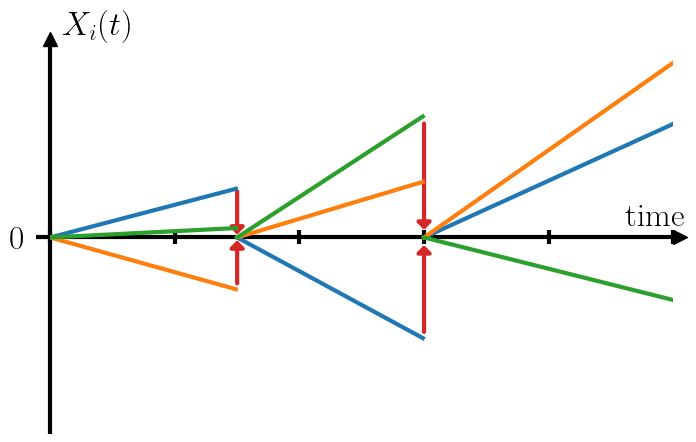}
\end{minipage}\hfill
\begin{minipage}[b]{0.33\textwidth}
\centering
\includegraphics[width = \textwidth]{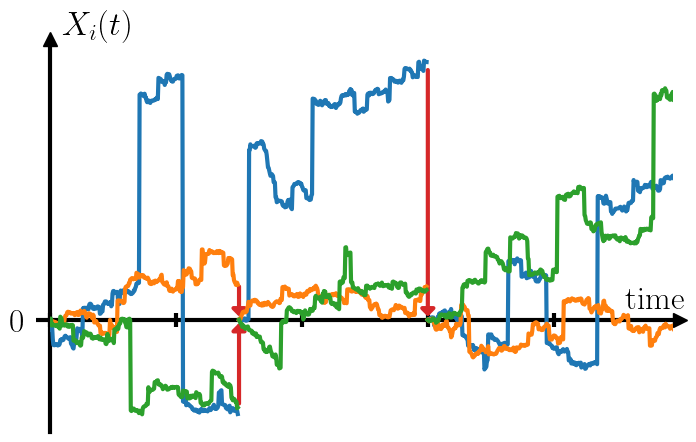}
\end{minipage}
\caption{A typical realization of three simultaneously resetting processes for $N=3$ particles (with different colors): (i) $N$ Brownian motions on a line, belonging to the Gumbel class of c.i.i.d. variables, (ii) $N$ ballistic particles on a line, with random velocities drawn from a uniform distribution over $[-1,+1]$, belonging to the Weibull class of c.i.i.d. variables and (iii) $N$ L\'evy flights on a line, belonging to the Fr\'echet class of c.i.i.d. variables. The red vertical arrows indicate the simultaneous resetting events.}
\label{Fig_intro}
\end{figure}

Once these general results are elucidated, it is natural to look for examples in physical systems where the c.i.i.d. variables arise naturally. As mentioned earlier, the case of $N$ Brownian motions on a line, all starting at the origin at $t=0$ and are simultaneously reset to the origin with rate $r$, provides a natural example of such c.i.i.d. variables~\cite{Biroli_23}. For a typical realisation of this process for $N=3$ particles, see the left panel of Fig. \ref{Fig_intro}. In this case, it was shown that the joint distribution of the positions approaches a non-equilibrium stationary state given by~\cite{Biroli_23} 
\begin{equation} \label{eq:reset_example}
{\rm Prob.}[X_1, \cdots, X_N] = \int_0^\infty \left\{\prod_{i = 1}^N \frac{1}{\sqrt{4 \pi D \tau}} e^{ - \frac{X_i^2}{4 D \tau}} \right\} \left(r e^{-r \tau} \right) \; \dd \tau \;. 
\end{equation}
A detailed derivation of this result is provided in Section \ref{brownian}, where we show that the EVS for this c.i.i.d. model can be computed exactly.
Note that without the averaging over $\tau$, i.e., for a fixed $\tau$, the $X_i$'s are i.i.d. variables, each distributed via a Gaussian. Hence for fixed $\tau$, their EVS converges to the Gumbel limiting distribution. Therefore this example provides a generalisation of the Gumbel class to c.i.i.d. variables. It is natural to ask whether there are physical models of c.i.i.d. variables whose i.i.d. counterparts have EVS belonging to the other two universality classes, namely the Fr\'echet and the Weibull classes. In fact, we introduce two new models of $N$ independent particles on a line that precisely do this job. For the Weibull class, we consider a model of $N$ independent ballistic particles, all starting at the origin with random velocities $v_i$'s each drawn independently from a uniform distribution $v_i \in [-1,+1]$, that are simultaneously reset to the origin with rate $r$. After every resetting, the velocities are also renewed. For a typical realisation of this ballistic resetting process for $N=3$ particles, see the middle panel in Fig. \ref{Fig_intro}. We show that the EVS for this c.i.i.d. model can also be computed exactly. Finally, we study a c.i.i.d. model 
whose i.i.d. counterpart corresponds to the Fr\'echet class for the EVS. In this model, we consider $N$ independent L\'evy flights, all starting at the origin and are reset simultaneously to the origin with rate $r$. In the rightmost panel of Fig. \ref{Fig_intro}, a typical realization of this process for $N=3$ is shown. For this model also, we show that the EVS can be computed exactly, despite the fact that the system is strongly correlated.

The rest of the paper is organized as follows. 
In Section \ref{section:iid} we recall the well-known results for the i.i.d. variables. 
The c.i.i.d. variables are introduced in Section \ref{ciid}. \blue{The results for the rescaled sum of c.i.i.d. variables are given in Subsection \ref{Sum-res} while the derivations are detailed in the Appendix. The EVS for general
c.i.i.d. variables are studied Subsection \ref{Order}.} The three physical models are
introduced and analysed in Section \ref{Illustration}. Finally, we conclude in Section \ref{conclusion}.

\section{Known results for independent and identically distributed random variables} \label{section:iid}

\blue{The statistics of i.i.d. random variables have been widely studied and are fully understood. We recall in this section the known results for the order statistics and the extreme value statistics of i.i.d. variables \cite{Gnedenko,Arnold,Nagaraja,Schehr_14,Majumdar_20}.} We consider $N$ i.i.d. variables \blue{$X_1, \ldots, X_N$} each distributed via $p(X)$ and we denote their cumulative distribution function (c.d.f.) by $P(X) = {\rm Prob.}[X_i \leq X]$. We define the order statistics by sorting them in decreasing order of magnitude
\begin{equation}
X_{\rm max} = M_{1, N} \geq M_{2, N} > \cdots > M_{N, N} = X_{\rm min} \;,
\end{equation}
where $X_{\rm max}$ (resp. $X_{\rm min}$) denotes the maximum (resp. minimum) among the $X_i$'s. We define the c.d.f. of the $k$-th maximum 
\begin{equation} \label{def_Fk}
F_{k,N}(w) = {\rm Prob.}[M_{k,N} \leq w] \;.
\end{equation}
When $k = O(N)$, we will refer to $M_{k,N}$ as the ``bulk'' order statistics, while the case $k=O(1)$ will be referred to as the extremes at the ``edge''. In particular, the case $k=1$, i.e., the distribution of the global maximum $M_{1,N}$, will be referred to as the EVS.

We start by characterizing the maximum, i.e., $X_{\rm max} = M_{1,N}$ which corresponds to $k=1$. In this case, the c.d.f. (appropriately centered and scaled) converges to a limiting scaling form 
\begin{equation}
F_{1, N}(w)  \underset{N \to +\infty}{\longrightarrow}  G_\rho\left( \frac{w - a_N}{b_N} \right) \;.
\end{equation}
The scale factors $a_N$ and $b_N$ depend on the details of the p.d.f. $p(x)$. However, it turns out that the scaling function $G_\rho(z)$
can be only of three varieties labelled by $\rho = {\rm I, II, III}$ (respectively the Gumbel, the Fr\'echet and the Weibull). When $N \to \infty$,  the c.d.f. $F_{1, N}(w)$ converges to one of these three scaling functions, depending on the tail of $p(x)$ as discussed below. \\

\vspace*{0.3cm}

\noindent{\bf (I) The Gumbel universality class} \\
If the p.d.f. $p(X)$ decays faster than any power law for large $X$, i.e. when $X \gg 1$ then $p(X) \ll X^{-\eta}$ for any $\eta > 0$ then $p(X)$ is said to belong to the Gumbel universality class. 
In this case
\begin{equation} \label{eq:Gumb}
G_{\rm I}(z) = e^{-e^{-z}}, \quad \int_{a_N}^{+\infty} p(X) \dd X = \frac{1}{N} \mbox{~~and~~} b_N = N \int_{a_N}^{+\infty} (x - a_N)p(X) \dd X\;.
\end{equation}\\
The scaling factors $a_N$ and $b_N$ are fixed by the last two equations in \ref{eq:Gumb}.

\vspace*{0.5cm}
\noindent{\bf (II) The Fr\'echet universality class} \\
If the p.d.f. $p(X)$ has an unbounded support and decays as a power law for large $X$, i.e. when $X \gg 1$ then $p(X) \propto X^{-1-\mu}$ for a given $\mu > 0$, then $p(X)$ is said to belong to the Fr\'echet universality class. 
In this case 
\begin{equation} \label{eq:props_frechet}
G_{\rm II}(z) = \Theta(z) e^{-z^{-\mu}}, \quad a_N = 0 \mbox{~~and~~} \int_{b_N}^{+\infty} p(X) \dd X = \frac{1}{N}\;,
\end{equation}
where $\Theta(z)$ is the Heaviside distribution.\\

\noindent{\bf (III) The Weibull universality class} \\
If the p.d.f. $p(X)$ has a support bounded from above at $X^\star$ and decays as a power law when approaching $X^\star$, i.e. when $X^\star - X \ll 1$ then $p(X) \propto (X^\star - X)^{\mu - 1}$ for a given $\mu > 0$, then $p(X)$ is said to belong to the Weibull universality class. 
In this case
\begin{equation}
G_{\rm III}(z) = \begin{cases}
1, \quad &z > 0\\
e^{-|z|^\mu}, &z < 0
\end{cases}
, \quad a_N = X^\star \mbox{~~and~~} \int_{X^\star - b_N}^{X^\star} p(X) \dd X = \frac{1}{N}\;.
\end{equation}
With these three classes we have universally characterized the EVS of i.i.d. variables. \\

It turns out that the c.d.f. of the $k$-th maximum $M_{k,N}$ when $k = O(1)$ can also be expressed in terms of the three scaling functions $G_\rho(z)$
 \begin{equation} \label{iid_os} 
{\rm Prob.}(M_{k,N} \leq w)  \underset{N \to +\infty}{\longrightarrow}  {\cal F}_{\rho}^{k}\left( \frac{w-a_N}{b_N}\right) \;,
\end{equation}
with the scaling function 
\begin{equation} \label{Fk_rho}
{\cal F}_{\rho}^{k}(z) =  \frac{1}{\Gamma(k)} \int_{ - \log G_\rho(z) }^{+\infty} e^{-t} \, t^{k-1} \dd t, \quad \rho = {\rm I, II, III} \;,
\end{equation}
where $\Gamma(z)$ is the Gamma function. For $k=1$, one can easily check that ${\cal F}_{\rho}^{k=1}(z) = G_\rho(z)$. When $k=O(N)$, i.e., one is probing the order statistics $M_{k,N}$ in the ``bulk'', the large $N$ limiting form of the p.d.f. of $M_{k,N}$ is known to be a Gaussian~\cite{Arnold,Nagaraja}  
\begin{equation} \label{eq:Mk_gauss_intro}
\text{Prob.}\,[M_{k = \alpha N, N} = w] \underset{N \to +\infty}{\longrightarrow}  \sqrt{\frac{N p[q(\alpha)]^2 }{2 \pi \alpha (1 - \alpha)}} \exp\left( - \frac{N p[q(\alpha)]^2}{2 \alpha(1 - \alpha)} [w - q(\alpha)]^2 \right) \;.
\end{equation}
where $0< \alpha < 1$ and $q(\alpha)$ is the $\alpha$-quantile obtained implicitly from
\begin{equation} \label{eq:def_quantile}
\int_{q(\alpha)}^{+\infty} p(X) \; \dd X = \alpha \;.
\end{equation} 
The quantile $q(\alpha)$ denotes the value of $X_i$ such that there are, on an average, $\alpha N$ variables above $q(\alpha)$.

\section{Conditionally independent and identically distributed random variables} \label{ciid}

We recall from Eq. \ref{def:general} in the introduction that the joint distribution of $N$ c.i.i.d. variables conditioned on $\vec{Y} = \{Y_1, Y_2, \cdots, Y_M\}$ is given by
\begin{equation}\label{def:general2}
{\rm Prob.}[X_1, \cdots, X_N] = \int \left\{\prod_{i = 1}^N p(X_i | \vec{Y})\right\} h(\vec{Y}) ~\dd Y_1 \ldots \dd Y_M\;,
\end{equation}
where $p(X_i | \vec{Y}) = {\rm Prob.}(X_i| Y_1, Y_2, \cdots, Y_M)$ is the conditional p.d.f. of $X_i$ and $h(\vec{Y})$ denotes the joint distribution of $\vec{Y} = \{Y_1, Y_2, \cdots, Y_M \}$. It is clear from Eq. \ref{def:general2} that the joint p.d.f. of $X_i$'s is not factorizable in general. Hence the $X_i$'s are correlated. By choosing $h(\vec{Y})$, one can generate a wide class of such correlated variables. To analyse the nature of the correlations, it is convenient to compute the connected correlator $\langle X_i^n X_j^n \rangle - \langle X_i^n \rangle \langle X_j^n \rangle$, for a generic $n$. To compute this, we first need the $n$-th moment $\langle X_i^n \rangle$, which is simply
\begin{equation} \label{eq:Xin}
\langle X_i^n \rangle = \int \dd \vec{X} \; X_i^n \; {\rm Prob.} [\vec{X}] = \int \dd \vec{X} \; X_i^n \int \dd \vec{Y} \; h(\vec{Y}) \prod_{k = 1}^N p(X_k | \vec{Y}) \;,
\end{equation}
where $\vec{X} = \{X_1, X_2, \cdots, X_N \}$ and $\dd \vec{X} = \dd X_1 \dd X_2 \cdots \dd X_N$. Consequently, the connected correlator is given by
\begin{equation} \label{connected}
\langle X_i^n X_j^n \rangle - \langle X_i^n \rangle \langle X_j^n \rangle = \int \dd \vec{Y}\; h(\vec{Y})  \int \dd \vec{X} X_i^n X_j^n \prod_{k = 1}^N p(X_k | \vec{Y})   -  \langle X_i^n \rangle \langle X_j^n \rangle   \;,
\end{equation}
where $\langle X_i^n \rangle $ is given in Eq. \ref{eq:Xin}. For a generic $h(\vec{Y})$, this connected correlator is nonzero, indicating
the presence of all-to-all correlations between the $X_i$ variables, which thus make them strongly correlated. 

\blue{\subsection{The scaled sum of c.i.i.d. variables}}\label{Sum-res}

\blue{
The main focus of this paper is on the order and the extreme statistics for c.i.i.d. variables that we discuss in detail in the later sections. However, it is also of interest to compute the statistics of the sum of such c.i.i.d. variables and to explore if there is an equivalent to the central limit theorem or L\'evy stable theorem for the c.i.i.d. case. In order not to shift the focus away from the order and the extreme statistics, in this section we present only the main results for the statistics of the sum, and provide the detailed derivations in the Appendix.}

\blue{Let us now summarize the main results for the rescaled sum $C = \frac{1}{N} \sum_{i = 1}^N X_i$ of c.i.i.d. variables $X_1, \cdots, X_N$. We defer the derivation of these results to the Appendix. If the conditional distribution $p(X_i | \vec{Y})$ admits a finite first moment $m(\vec{Y})$ which is a non-constant function of $\vec{Y}$ and a finite second moment $\sigma(\vec{Y})$ then the p.d.f. $P(C, N)$ of the rescaled sum can be written as
\begin{equation} \label{CLT-m-res}
P(C, N) \stackrel{N \to +\infty}{\longrightarrow} \int \dd \vec{Y} \; \delta(m(\vec{Y}) - C) \; h(\vec{Y}) \;.
\end{equation}
On the other hand if $m(\vec{Y}) = m$ is a constant then
\begin{equation} \label{CLT-var-res}
P(C, N) \stackrel{N \to +\infty}{\longrightarrow} \sqrt{N} \mathcal{P}\Big( (C - m) \sqrt{N} \Big) \mbox{~~where~~} \mathcal{P}(Z) = \frac{1}{\sqrt{2 \pi}} \int \frac{\dd \vec{Y}}{\sigma(\vec{Y})} \exp( - \frac{Z^2}{2 \sigma^2 (\vec{Y})}) h(\vec{Y}) \;.
\end{equation}
Finally, if $p(X_i | \vec{Y})$ does not admit a finite first or second moment and instead has a power law tail $p(X_i | \vec{Y}) \sim 1/X^{1 + \mu}$ for large $X$, with $0 < \mu < 2$ then
\begin{equation} \label{CLT-L-res}
P(C, N) \approx N^{1 - 1/\mu} \mathcal{P}_\mu\left( \frac{C}{N^{1/\mu - 1}} \right) \mbox{~~where~~} \mathcal{P}_\mu(Z) = \int \frac{\dd \vec{Y}}{b(\vec{Y})} \mathcal{L}_\mu \left( \frac{Z}{b(\vec{Y})} \right) h(\vec{Y}) \;,
\end{equation}
where ${\cal L}_{\mu}(z)$ is the L\'evy stable distribution (scaled to unity) \cite{Bouchaud-Georges}
\begin{eqnarray} \label{stable}
{\cal L}_{\mu}(z) = \int_{-\infty}^\infty \frac{\dd q}{2 \pi} e^{- i q z - |q|^\mu} \;.
\end{eqnarray}
It has the asymptotic behaviors \cite{Bouchaud-Georges}
\begin{equation} \label{eq:tail_stable_law_intro}
\mathcal{L}_\mu(z) 
\approx
\begin{cases}
& \frac{1}{\pi \mu}  \Gamma(1/\mu) \quad, \quad \quad \quad \; \; z \to 0 \;,\\
& \\
&\frac{1}{z^{1 + \mu}} \sin(\frac{\pi \mu}{2}) \frac{\Gamma(\mu + 1)}{\pi} \quad z \to \infty \;.
\end{cases}
\end{equation}}

\subsection{Order statistics of c.i.i.d. variables}\label{Order}

In this section, we provide a complete characterization of the order statistics and EVS for c.i.i.d. variables. 
We consider again a set of $N$  c.i.i.d. random variables $X_1, \cdots, X_N$. Since the $X_i$'s are conditionally independent, the p.d.f. of the $k$-th maxima $M_{k, N}$ conditioned on $\vec{Y}$ is given by
\begin{equation} \label{eq:exact_Mk}
\text{Prob.}[M_{k, N} = w | \vec{Y}] = \frac{N!}{(k-1)!(N - k)!} p(w | \vec{Y}) \left[ \int_{w}^{+\infty} p( X | \vec{Y}) \dd X \right]^{k-1} \left[ \int_{-\infty}^{w} p(X| \vec{Y}) \dd X \right]^{N- k} \;.
\end{equation}
This equation is exact for any $k$ and any distribution $p( X | \vec{Y})$ and can be understood as follows. 
For the $k$-th maximum to be located at $w$, we need to place one variable exactly at $w$, $k-1$ variables above $w$ and $N-k$ variables 
below $w$. The p.d.f. of the $k$-th maximum, integrated over $\vec{Y}$, reads
\begin{equation} \label{eq:total_prob}
\text{Prob.}[M_{k, N} = w] = \int \dd \vec{Y}\; h(\vec{Y}) \; \text{Prob.}[M_{k, N} = w | \vec{Y}] \;.
\end{equation}
We now study the large $N$ limit of this p.d.f. by setting $k = \alpha N$ where $0 < \alpha < 1$. Eq. \ref{eq:exact_Mk} then gives
\begin{equation} \label{eq:Malpha}
{\rm Prob.}[M_{k, N} = w | \vec{Y}] = \frac{\Gamma(N+1)}{\Gamma(\alpha N) \Gamma(N(1 - \alpha) + 1)} \frac{p( w | \vec{Y})}{\int_{w}^{+\infty} p( X | \vec{Y}) \dd X} e^{- N \Phi_\alpha(w)} \;,
\end{equation}
where we re-wrote the combinatorial factor using the Gamma function $\Gamma(z)$ and we introduced the function 
\begin{equation} \label{eq:def_phi}
\Phi_\alpha(w) = -\alpha \ln\left( \int_{w}^{+\infty} p( X | \vec{Y}) \dd X \right) - (1 - \alpha) \ln \left( \int_{-\infty}^{w} p( X | \vec{Y}) \dd X \right) \;.
\end{equation}
So far, we have not taken the large $N$ limit. When $N \to \infty$, the p.d.f. in \ref{eq:Malpha} gets sharply concentrated around the value $w = q(\alpha, \vec{Y})$ that minimises $\Phi_\alpha(w)$. Setting $\Phi_\alpha'(w = q(\alpha, \vec{Y})) = 0$ yields 
\begin{equation} \label{eq:def_wstar}
\int_{q(\alpha, \vec{Y})}^{+\infty} p(X | \vec{Y}) \; \dd X = \alpha \;.
\end{equation} 
Notice that this corresponds to the definition of the $\alpha$-quantile of the conditional distribution $p(X | \vec{Y})$. 
Then expanding $\Phi_\alpha(w)$ around $q(\alpha, \vec{Y})$ up to quadratic order, one finds that for large $N$ and close to $q(\alpha, \vec{Y})$, the p.d.f. defined in Eq. \ref{eq:Malpha} simplifies to
\begin{equation} \label{eq:Mk_gauss}
\text{Prob.}[M_{k, N} = w | \vec{Y}] \approx \sqrt{\frac{N \left[p (q \,|\,  \vec{Y})\right]^2 }{2 \pi \alpha (1 - \alpha)}} \exp\left( - \frac{N \left[p(q \,|\, \vec{Y})\right]^2}{2 \alpha(1 - \alpha)} [w - q]^2 \right) \;,
\end{equation}
where for brevity we suppressed the explicit dependence on $\alpha$ and $\vec{Y}$ of $q \equiv q(\alpha, \vec{Y})$. Then Eq. \ref{eq:Mk_gauss} is simply a Gaussian distribution centered around $q$ with variance 
\begin{equation} \label{eq:variance_main}
{\rm Var} = \frac{\alpha(1-\alpha)}{N \left[p(q \, | \, \vec{Y})\right]^2} \;. 
\end{equation}
For the order statistics in the bulk where $\alpha \sim \mathcal{O}(1)$, it follows from Eq. \ref{eq:def_wstar} that $q(\alpha, \vec{Y})$ is also of order $\mathcal{O}(1)$. Consequently, from Eq. \ref{eq:variance_main}, one finds that ${\rm Var} \sim \mathcal{O}(1/N)$. Under these conditions, in the large $N$ limit, the p.d.f. in Eq. \ref{eq:Mk_gauss} becomes sharply peaked and can be approximated by a Dirac delta function centered at $q(\alpha, \vec{Y})$. 
Note that this is true regardless of the underlying distribution $p(X | \vec{Y})$. Substituting the Dirac delta function back in Eq. \ref{eq:total_prob}, we find that the p.d.f. of the $k$-th maximum in the bulk converges, in the large $N$ limit, to an $N$-independent limiting form given by
\begin{equation} \label{eq:res_bulk_integral_form}
\text{Prob.}[M_{k, N} = w] \underset{N \to \infty}{\longrightarrow} \int \dd \vec{Y} \; \; h(\vec{Y}) \, \delta[w - q(\alpha, \vec{Y})] \;.
\end{equation}
This result characterizes the order statistics in the bulk for c.i.i.d. variables. 

However, for the order statistics near the edge, where $\alpha = \mathcal{O}(1/N)$, we need to carefully study the dependence on $N$ of the quantile $q(\alpha, \vec{Y})$ and the variance~${\rm Var}$. From Eq. \ref{eq:def_wstar} we see that when $\alpha \sim \mathcal{O}(1/N) \ll 1$, the quantile $q(\alpha, \vec{Y})$ depends on the tail of $p(X | \vec{Y})$ for large $X$. We therefore analyse separately the three classes of tails that lead to the three universality classes of the EVS in the i.i.d. case. In section \ref{gumbel}, we study the Gumbel class, where the distribution $p(X \, | \vec{Y})$ decays faster than a power law for large $X$. 
In section \ref{weibull}, we study the Weibull class, where the support of the distribution $p(X | \vec{Y})$ is bounded above and approaches its upper-bound as a power-law. Finally, in section \ref{frechet}, we study the Fr\'echet class, where the distribution $p(X  | \vec{Y})$ has an unbounded support and decays as a power-law for large $X$. We will see that while the bulk result in Eq. \ref{eq:res_bulk_integral_form} can be extrapolated to the edge (where $k = \mathcal{O}(1)$) for the Gumbel and the Weibull class, the same can not be done for the Fr\'echet case. In the Fr\'echet case, the order statistics at the edge needs to be analysed separately. 

\subsubsection{Edge order statistics for the Gumbel class}\label{gumbel}

We start by studying random variables belonging to the Gumbel class, e.g., when the tail of the distribution $p(X_i | \vec{Y})$ decays as 
\begin{equation} \label{eq:gumbel_tail}
p(\blue{X_i} | \vec{Y}) \underset{\blue{X_i} \gg 1}{\sim}  A(\vec{Y}) e^{- B(\vec{Y}) \blue{X_i}^{\mu(\vec{Y})} } \;,
\end{equation}
where $A(\vec{Y}), B(\vec{Y}), \mu(\vec{Y}) > 0$ are positive real functions of $\vec{Y}$.
For simplicity, we will drop the explicit dependence on $\vec{Y}$ of $A(\vec{Y}), B(\vec{Y})$ and $\mu(\vec{Y})$ and simply denote them by $A, B$ and $\mu$. 
In order to probe the behavior of the extremes near the edge, i.e., of $M_{k,N}$ for $k \sim \mathcal{O}(1)$. To do so, we start by studying the dependence on $N$ of the quantile $q(\alpha, \vec{Y})$ when $\alpha = k/N = \mathcal{O}(1/N)$. From Eq. \ref{eq:def_wstar} we see that when $\alpha \ll 1$ then necessarily $q(\alpha, \vec{Y}) \gg 1$. Hence, in Eq. \ref{eq:def_wstar}, we can replace the integrand by its tail behavior \ref{eq:gumbel_tail} and obtain
\blue{\begin{equation} \label{eq:compute_quantile_gumbel}
\alpha \approx \int_{q(\alpha, \vec{Y})}^{+\infty}  \; A \, e^{-B x^{\mu}} \; \dd x = \frac{A}{\mu B^{1/\mu}} \; \int_{B \, q^\mu(\alpha, \vec{Y}) }^{+\infty} \; t^{1/\mu - 1} \, e^{-t} \; \dd t \; \; \underset{q(\alpha, \vec{Y}) \gg 1}{\approx} \;\;  
 \frac{A}{B \, \mu} \, q^{1 - \mu}(\alpha, \vec{Y})\, e^{-B \, q^{\mu}(\alpha, \vec{Y})}  \;\blue{.}
\end{equation}}
\blue{where we approximated the integral by it's value at it's lower bound which is the leading order contribution due to the exponential decay of the integrand.} Since we are at the edge, where $\alpha = k/N \ll 1$ and $q(\alpha, \vec{Y}) \gg 1$, the leading order solution for $q(\alpha, \vec{Y})$ from Eq. \ref{eq:compute_quantile_gumbel} is given by 
\begin{equation} \label{eq:qaY_gumbel_tail}
q(\alpha, \vec{Y}) \underset{\alpha \to 0}{ \approx} \frac{1}{B^{1/\mu}} \blue{\log^{1/\mu}}\left( \frac{A}{B \alpha \mu} \right) \;.
\end{equation}
Replacing $\alpha = k/N$ with $k \sim \mathcal{O}(1)$ fixed and taking $N \to +\infty$ we obtain
\begin{equation} \label{q_alpha_gumbel}
q(\alpha, \vec{Y}) \approx \frac{1}{B^{1/\mu}} \blue{\log^{1/\mu}}\left( N \frac{A}{B  k \mu} \right) \sim \mathcal{O}\left[ \blue{\log^{1/\mu}}(N) \right] \;,
\end{equation}
thus justifying a posteriori that $q(\alpha, \vec{Y}) \gg 1$ for large $N$. We now investigate how the variance ${\rm Var}$ in Eq. \ref{eq:variance_main} 
depends on $N$ for large $N$. To do so, we substitute Eq. \ref{q_alpha_gumbel} back into Eq. \ref{eq:gumbel_tail} yielding
\begin{equation} \label{eq:pqY_gumbel_tail}
p(  q \, | \, \vec{Y} ) \underset{\alpha \ll 1}{ \approx} \frac{B k \mu}{N} \;.
\end{equation}
Using Eq. \ref{eq:pqY_gumbel_tail} in Eq. \ref{eq:variance_main}, we get that the variance is given by
\begin{equation}
{\rm Var} = \frac{\alpha (1 - \alpha)}{N \left[  p(q\,  | \, \vec{Y}) \right]^2 }  \approx \frac{k/N}{N \left( \frac{B k \mu}{N} \right)^2}  \approx \frac{1}{\blue{B^2} \mu^2 k} \sim \mathcal{O}(1)\;.
\end{equation}
Thus, the variance ${\rm Var} = \mathcal{O}(1)$ while from Eq. \ref{q_alpha_gumbel} the mean $q(\alpha, \vec{Y}) = \mathcal{O}( [\log N]^{1/\mu} )$. Hence, we can write
\begin{equation}\label{def_etak}
M_{k, N} \approx q(\alpha, \vec{Y})+ \eta_k\;,
\end{equation}
where $\eta_k$ is a random variable of order $\mathcal{O}(1)$. Therefore, in the large $N$ limit, the fluctuations are negligible compared to the mean for any positive $A(\vec{Y}), B(\vec{Y})$ and $\mu(\vec{Y})$. For large $N$, the $k$-th maximum $M_{k, N}$ effectively concentrates on its mean value $M_{k, N} \approx q(\alpha, \vec{Y}) \sim \blue{\log^{1/\mu}}(N)$.
Thus, the approximation used to obtain Eq. \ref{eq:res_bulk_integral_form} is still valid even at the edge, i.e. for $\alpha \sim \mathcal{O}(1/N)$. Hence, for the Gumbel class, Eq. \ref{eq:res_bulk_integral_form} characterizes both the order statistics in the bulk as well as at the edge for c.i.i.d. variables.

\subsubsection{Edge order statistics for the Weibull class}\label{weibull}

We now turn our attention to c.i.i.d. variables belonging to the Weibull class. In this case, the support of the distribution $p(\blue{X_i} | \vec{Y})$ is bounded above by $x^\star(\vec{Y})$, and approaches this bound with a power-law tail given by
\begin{equation} \label{tail_weibull}
p(\blue{X_i} | \vec{Y}) \blue{\underset{\blue{X_i} \to x^\star(\vec{Y})}{\approx}}  A(\vec{Y}) (x^\star(\vec{Y})- \blue{X_i})^{\mu(\vec{Y}) - 1} \;,
\end{equation}
for $\blue{X_i}$ close to $x^\star(\vec{Y})$, where $A(\vec{Y}), x^\star(\vec{Y}), \mu(\vec{Y}) > 0$ are positive real functions of $\vec{Y}$, where for simplicity we will drop the explicit dependence on $\vec{Y}$.
We start by studying the dependence on $N$ of the quantile $q(\alpha, \vec{Y})$. Once again, to probe the EVS we have to fix $k \sim \mathcal{O}(1)$ while taking $N \to +\infty$ to study the behavior close to the global maximum $M_{1,N}$. 
Hence $\alpha\sim\mathcal{O}(1/N)\ll 1$ and $q(\alpha, \vec{Y})$ must be approaching $x^\star$ from below. 
This justifies using the tail expression given in Eq. \ref{tail_weibull} inside the integral in Eq. \ref{eq:def_wstar}. Hence,
\begin{equation}
\alpha = \int_{q(\alpha, \vec{Y})}^{+\infty} p(X | \vec{Y}) \; \dd X = A \int_{q(\alpha, \vec{Y})}^{x^\star} (x^\star - X)^{\mu - 1} \dd X = \frac{A}{\mu} \, (x^\star - q(\alpha, \vec{Y}))^{\mu} \;,
\end{equation}
which gives
\begin{equation} \label{q_alpha_weibull}
q(\alpha, \vec{Y}) = x^\star - \left(\frac{\alpha \mu}{A}\right)^{1/\mu} \;.
\end{equation}
In the limit of $N\to+\infty$, keeping $k \sim \mathcal{O}(1)$ fixed, $\alpha = k/N \to 0$ and therefore $q(\alpha, \vec{Y}) \sim \mathcal{O}(1)$. We now study the dependence on $N$ of the variance ${\rm Var}$. Plugging Eq. \ref{q_alpha_weibull} back into Eq. \ref{tail_weibull} we get
\begin{equation}
p(q \, | \, \vec{Y}) \approx A^{1/\mu} (\alpha \mu)^{1 - 1/\mu} \;.
\end{equation}
Thus, the variance in Eq. \ref{eq:variance_main}, is given by
\begin{equation} \label{var_alpha_weibull}
{\rm Var} = \frac{\alpha (1 - \alpha)}{ N \left[p(q \,|\, \vec{Y})\right]^2 } \approx \frac{k/N (1 - k/N)}{ N A^{2\mu} (k \mu/N)^{2 - 2/\mu} } \approx \frac{k}{A^{2\mu} (k \mu)^{2 - 2/\mu} } N^{-2\mu} \sim \mathcal{O}(N^{-2\mu})  \;,
\end{equation}
where we used $\alpha = k/N$ and took the EVS limit of $N \to +\infty$ keeping $k = \mathcal{O}(1)$ fixed.
From Eqs. \ref{q_alpha_weibull} and \ref{var_alpha_weibull} we see that the mean is of order $\mathcal{O}(1)$ while the variance is of order $\mathcal{O}(N^{-2\mu})$.
Hence, the approximation that the Gaussian is sharply peaked holds and the result derived in Eq. \ref{eq:res_bulk_integral_form} is valid for any $k$. 
Hence, for the Weibull class, Eq. \ref{eq:res_bulk_integral_form} also fully characterizes both the order statistics in the bulk as well as at the edge for c.i.i.d. random variables.

\subsubsection{Edge order statistics for the Fr\'echet class}\label{frechet}

Finally, we turn our attention to c.i.i.d. variables belonging to the Fr\'echet class. The conditional distribution $p(\blue{X_i}  | \vec{Y})$ has an unbounded support with a power-law tail given by
\begin{equation} \label{eq:tail_frechet}
p(\blue{X_i} | \vec{Y}) \underset{\blue{X_i} \gg 1}{\approx}  \frac{A(\vec{Y})}{\blue{X_i}^{1 + \mu(\vec{Y})}} \;,
\end{equation}
where $A(\vec{Y}) > 0$ and $2 > \mu(\vec{Y}) > 0$ are positive real functions of $\vec{Y}$.
Once more, we drop the explicit dependence on $\vec{Y}$ of $A(\vec{Y})$ and $\mu(\vec{Y})$ for brevity and we will restore it later. 
To probe the EVS, we have to take the limit of $N \to +\infty$ keeping $k = \mathcal{O}(1)$ fixed to study the behavior close to the global maximum $M_{1,N}$.
In this limit, $\alpha = k/N = \mathcal{O}(1/N) \ll 1$, which from Eq. \ref{eq:def_wstar} implies $q(\alpha, \vec{Y}) \gg 1$.
Therefore, the integrand in Eq. \ref{eq:def_wstar} can be replaced by its tail behaviour given in Eq. \ref{eq:tail_frechet}. Then,
\begin{equation}
\alpha \approx \int_{q(\alpha, \vec{Y})}^{+\infty} A \frac{\dd X}{X^{1 + \mu}} \approx \frac{A}{\mu} \blue{q^{-\mu}}(\alpha, \vec{Y}) \;,
\end{equation}
hence
\begin{equation}
q(\alpha, \vec{Y}) \approx \left(\frac{A}{\mu \alpha}\right)^{1/\mu} \;.
\end{equation}
Taking, $\alpha = k/N$ and the limit $N \to +\infty$ keeping $k = \mathcal{O}(1)$ fixed results in
\begin{equation} \label{eq:q_alpha_frechet}
q(\alpha, \vec{Y}) \approx \left(\frac{A N}{\mu k}\right)^{1/\mu}  \sim \mathcal{O}(N^{1/\mu})\;.
\end{equation}
This characterizes the dependence on $N$ of the quantile $q(\alpha, \vec{Y})$. We now turn our attention to the dependence on $N$ of the variance ${\rm Var}$. 
Plugging Eq. \ref{eq:q_alpha_frechet} back in Eq. \ref{eq:tail_frechet} we get
\begin{equation}
p( q \,|\, \vec{Y}) \underset{\alpha \ll 1}{\approx}  A^{1/\mu} \left(\alpha\right)^{1/\mu + 1} \approx \frac{A^{1/\mu} k^{1/\mu + 1}}{N^{1/\mu + 1}} ,
\end{equation}
where we used $\alpha = k/N$. Then, the variance in Eq. \ref{eq:variance_main} is given by
\begin{equation}
{\rm Var} = \frac{\alpha(1- \alpha)}{N \left[p(q \, | \, \vec{Y})\right]^2} \sim  \mathcal{O}(N^{2/\mu}) \;.
\end{equation}
Thus the width of the fluctuations is of the same order as the mean value for large $N$, since ${\rm Var}/q(\alpha, \vec{Y})^2 \sim \mathcal{O}(1)$ and hence
the distribution does not concentrate on its mean for large $N$ -- at variance with the Gumbel and the Weibull class. 
Thus in the Fr\'echet case, we need to analyse the extremes at the edge directly from the limiting behaviour given in Eq. \ref{iid_os}, instead of extrapolating the bulk result to the edge. We recall the result in Eq. \ref{iid_os} for $\rho = {\rm II}$ corresponding to the Fr\'echet class. It reads, for $k \sim O(1)$, 
\begin{equation}
{\rm Prob.}[M_{k, N} \leq w \,|\, \vec{Y}] \approx \mathcal{F}^{k}_{{\rm II}}\left( \frac{w - a_N}{b_N} \right) \;,
\end{equation}
where $\mathcal{F}^{k}_{{\rm II}}(z)$ is given in Eq. \ref{Fk_rho}. From Eq. \ref{eq:props_frechet}, we know that the scale factors are given by
\begin{equation} \label{eq:aN_bN_frechet}
a_N = 0 \mbox{~~and~~} \int_{b_N}^{+\infty} p(X | \vec{Y}) \dd X = \frac{1}{N} \;.
\end{equation}
Comparing Eq. \ref{eq:aN_bN_frechet} with Eq. \ref{eq:def_wstar} we see immediately that 
\begin{equation}
b_N = q(1/N, \vec{Y}) \equiv q_N(\vec{Y}) \;.
\end{equation}
Putting the scale factors obtained in Eq. \ref{eq:aN_bN_frechet} in Eq. \ref{iid_os} we get
\begin{equation} \label{eq:Mk_2_Gk}
{\rm Prob.}[M_{k, N} \leq w \,|\, \vec{Y} ] \underset{N \to +\infty}{\longrightarrow} \frac{1}{\Gamma(k)} \int_{ w^{-\mu} q_N(\vec{Y})^\mu }^{+\infty} \dd t \; e^{-t} t^{k-1} \;.
\end{equation} 
where we used $G_{\rm II}(z) = \Theta(z)\,e^{-z^{-\mu}}$ from Eq. \ref{eq:props_frechet}.

Once again, in the $N \gg 1$ limit, we have $q_N(\vec{Y}) \gg 1$, so we can plug the tail behavior given in Eq. \ref{eq:tail_frechet} into Eq. \ref{eq:aN_bN_frechet} and solve for $q_N(\vec{Y})$, as was done before for $q(\alpha, \vec{Y})$. 
This yields
\begin{equation} \label{eq:bny}
q_N(\vec{Y}) \simeq \left( \frac{A N}{\mu} \right)^{1/\mu} \;.
\end{equation}
Now, using Eq. \ref{eq:Mk_2_Gk} in Eq. \ref{eq:total_prob}, we get
\begin{equation} \label{eq:intermed_1}
{\rm Prob.}[M_{k, N} \leq w]  = \int \dd \vec{Y} \; h(\vec{Y}) {\rm Prob.}[M_{k, N} \leq w \, | \, \vec{Y}] =  \frac{1}{\Gamma(k)} \int \dd \vec{Y} \; h(\vec{Y}) \int_{w^{-\mu} q_N(\vec{Y})^\mu}^{+\infty} \dd t \; e^{-t}t^{k-1} \;.
\end{equation}
We denote the lower bound of the integral over $t$ by 
\begin{equation} \label{eq:def_lambda}
\lambda_N(w, \vec{Y}) = \left(\frac{q_N(\vec{Y})}{w}\right)^{\mu(\vec{Y})} =  \frac{A(\vec{Y}) N}{ \mu(\vec{Y}) w^{\mu(\vec{Y})} }  \;,
\end{equation}
where we restored the explicit dependence of $A$ and $\mu$ on $\vec{Y}$. Next we express the integral over $t$ as
\bea \label{decomp}
\int_{\lambda_N(w, \vec{Y})}^{+\infty} \dd t \; e^{-t}\, t^{k-1} = \Gamma(k) - \int_0^{\lambda_N(w, \vec{Y})} \dd t \; e^{-t}t^{k-1}  = \Gamma(k) - \int_0^{\infty} \, \dd t \, \Theta\left(\lambda_N(w,\vec{Y})-t\right) \,  e^{-t}\,t^{k-1} \;.
\eea
Substituting this integral in Eq. \ref{eq:intermed_1} and using the fact that $h(\vec{Y})$ is normalized to unity, we get, 
\begin{eqnarray} \label{eq:intermed_2}
{\rm Prob.}[M_{k, N} \leq w] = 1 - \frac{1}{\Gamma(k)} \int_0^\infty \dd t \, e^{-t}\,t^{k-1} \int \dd \vec{Y} \, h(\vec{Y}) \,  \Theta\left(\lambda_N(w,\vec{Y})-t\right) \;.
\end{eqnarray}
Thus interpreting $\lambda_N(w, \vec{Y})$ as a random variable, we get
\begin{equation}\label{eq:res:frechet}
{\rm Prob.}[M_{k, N} \leq w] = 1 - \frac{1}{\Gamma(k)} \int_0^{\infty} \dd t \; {\rm Prob.}\left[\lambda_N(w, \vec{Y}) \geq t\right]\; e^{-t} \, t^{k-1} \;,
\end{equation}
where 
\bea \label{dist_lambda}
{\rm Prob.}\left[\lambda_N(w, \vec{Y}) \geq t\right] = \int \dd \vec{Y} \, h(\vec{Y}) \,  \Theta\left(\lambda_N(w,\vec{Y})-t\right) \;.
\eea
Given the distribution $h(\vec{Y})$ of $\vec{Y}$, it follows that $A(\vec{Y})$ and $\mu(\vec{Y})$ are random variables in Eq. \ref{eq:def_lambda}. We first need to compute the distribution of $\lambda_N(w,\vec{Y})$ defined in \ref{eq:def_lambda} using Eq. \ref{dist_lambda}. Finally, we need to substitute this c.d.f. of $\lambda_N(w,\vec{Y})$ in Eq. \ref{eq:res:frechet} to compute the c.d.f. of $M_{k,N}$. Hence Eq. \ref{eq:res:frechet} characterizes the order statistics at the edge for the  
c.i.i.d. variables belonging to the Fr\'echet class.  

\vspace{0.5cm}

\blue{
Let us end this section with the following remark. As discussed above we see that for c.i.i.d. variables belonging to the Gumbel or Weibull class (but not the Fr\'echet), one could obtain the statistics of the positions of the particles near the edges by extrapolating the bulk results. One may wonder if the same can be done for i.i.d. variables belonging to the Gumbel or the Weibull class. However, it is well known that for i.i.d. variables the bulk and the edge order statistics behave very differently and their statistics cannot be derived using just a single framework. Our results demonstrate that, in contrast to the i.i.d. case, a single function indeed describes the statistics of the particle positions both in the bulk and at the edges for c.i.i.d. variables belonging to the Gumbel and the Weibull class.
}

\section{Simultaneously resetting stochastic processes}\label{Illustration}

We now provide three concrete physical examples belonging to the three classes where the general results derived above for the c.i.i.d. variables can be applied. These examples are motivated by the recent advances in the field of stochastic resetting. Stochastic resetting simply means interrupting the natural dynamics of a system (deterministic or stochastic) at random times and restart from the same initial condition. 
The resetting breaks detailed balance and drives the system to a non-equilibrium stationary state (NESS). This NESS has been studied in various theoretical models~\cite{EM_11,EM_11b,KMSS_14,EM_14,MSS_15,Reu_16,MV_16,PKE_16,PR_17,BEM_17,CS_18,Bres_20,Pinsky_20,BMS_22,VAM_22,Bertin_22}, for recent reviews see \cite{EMS_20,Gupta22,PKR_22}. Some of these theoretical predictions have been verified in reent experiments on colloids diffusing in an optical trap \cite{TPSRR_20,BBPMC_20,FBPCM_21}. As stated in the introduction, a \blue{straightforward} and physically interesting way of obtaining c.i.i.d. variables is through simultaneous resetting of otherwise independent processes. It is a mechanism first introduced in \cite{Biroli_23}, which we generalize here to two new systems. 
Specifically, we consider $N$ independent particles whose positions $X_1(t), \cdots, X_N(t)$ (which we abbreviate as $\vec{X}(t)$) evolve on the line
under some dynamics which we refer to as its ``natural dynamics''. For example, it could be $N$ independent Brownian motions or $N$ independent ballistic particles, etc. For simplicity we consider the evolution in continuous time but this can easily be generalised to discrete time processes such as L\'evy flights. 
 We introduce resetting to the origin by
\begin{equation} \label{reset_dyn}
\vec{X}(t + \dd t) = \begin{dcases}
0, \cdots, 0 &\mbox{with probability~~} r \dd t \\
\mbox{each~} X_i \mbox{~evolves independently via its natural dynamics} &\mbox{with complementary probability~~} 1 - r \dd t
\end{dcases}\;.
\end{equation}
In other words, after every time step $\dd t$, we either reset all the processes simultaneously to the origin with probability $r \dd t$, or we let each process evolve independently, as it would have in the absence of resetting. A cartoon illustrating the trajectories for three particles evolving under this dynamics is shown in Fig. \ref{Fig_intro}. We denote by $p(X \,|\, \tau)$ the free-propagator (in the absence of resetting) at time $\tau$ of this process starting at $X=0$, i.e., the probability density for the process to arrive at $X$ at time $\tau$, starting at $X=0$.  
In the absence of resetting the variables $X_1(\tau), X_2(\tau), \cdots, X_N(\tau)$ are independent. Hence, for a given $\tau$, their joint distribution factorizes  
${\rm Prob.}[\vec{X} \, |\, \tau] = \prod_{i = 1}^N p(X_i \,|\, \tau)$. We can write a renewal equation for the simultaneously resetting process which reads
\begin{equation} \label{eq:general_renewal}
{\rm Prob.}[\vec{X}(t)] = e^{-r t} \prod_{i = 1}^N p(X_i \, |\, t)  + r \int_0^{t} \dd\tau \; e^{-r \tau} \prod_{i = 1}^N p(X_i | \tau) \;.
\end{equation}
This equation can be understood as follows. There is a possibility that the process never resets in the time interval $[0, t]$. 
The probability of never resetting in this interval is given by $e^{-r t}$ and the propagator $\prod_{i = 1}^N p(X_i \,|\, t)$ gives us the probability for the free-process to reach $\vec{X}$ at time $t$. 
This corresponds to the first term of Eq. \ref{eq:general_renewal}. 
Otherwise, the process will reset at least once before reaching $\vec{X}$ at time $t$. 
Suppose it has reset for the last time at time $t - \tau$, it then has to reach $\vec{X}$ from $\vec{0}$ in the time interval $[t - \tau, t]$ while never resetting. 
The probability of resetting once is given by $r \dd \tau$, the probability of never resetting in the interval $[t - \tau, t]$ is given by $e^{-r\tau}$ and the probability of reaching $\vec{X}$ from $\vec{0}$ in time $\tau$ is given by the free-propagator $\prod_{i = 1}^N p(X_i \,|\, \tau)$. 
Multiplying this free propagator by the distribution of $\tau$ and integrating over $\tau$ gives the second term of Eq. \ref{eq:general_renewal}. 
We can see from Eq. \ref{eq:general_renewal} that in the long time limit, $t \gg 1$, the first term drops out and the simultaneously resetting process reaches a NESS
\begin{equation} \label{eq:general_NESS}
{\rm Prob.}[\vec{X}]_{\rm NESS} = r \int_0^{+\infty} \dd\tau~ e^{-r \tau} \prod_{i = 1}^N p(X_i \,|\, \tau) \;.
\end{equation}
This steady state is out-of-equilibrium because resetting manifestly breaks detailed balance in the configuration space. 
Comparing Eq. \ref{def:general} and Eq. \ref{eq:general_NESS} we see that this system provides an example of c.i.i.d. variables. Here, we have only $M=1$ conditioning variable $Y_1 = \tau$, which is the time elapsed since the last resetting event before $t$. Since $\tau$ has an exponential distribution, we then have $h(\tau) = r e^{-r \tau}$ with parameter $r$. As discussed in the introduction, this joint probability density function does not factorize. 
Indeed the simultaneous resetting induces strong "all-to-all" correlations in our gas.

Given the joint distribution given in Eq. \ref{eq:general_NESS} in the NESS, one can investigate various physical observables, such as the average density of the gas, the order statistics, the statistics of gaps between particles or the full counting statistics, i.e., the distribution of the number of particles in a given interval \cite{Biroli_23}. For example, one of the basic observables is the average density of the gas given by
\begin{equation} \label{eq:gen_density}
\rho(x, N \vert r) = \left\langle \frac{1}{N} \sum_{i = 1}^N \delta(X_i - x) \right\rangle = r \int_0^{+\infty} \dd\tau\; e^{-r \tau} p(x \,|\, \tau) \;.
\end{equation}
Therefore the average density in the NESS is just the marginal distribution of a set of c.i.i.d. variables. Interestingly, the average density is independent of $N$. Similarly the order statistics of the positions of the particles in the NESS of this gas can be derived using the general results for c.i.i.d. variables detailed before. In this paper, we focus only on the average density and the order statistics, but in principle one can also obtain exactly other observables such as the gap and the full counting statistics.

\subsection{Brownian Motion: an example of the Gumbel class}
\label{brownian}

\begin{figure}
\centering
\includegraphics[width = 0.6\textwidth]{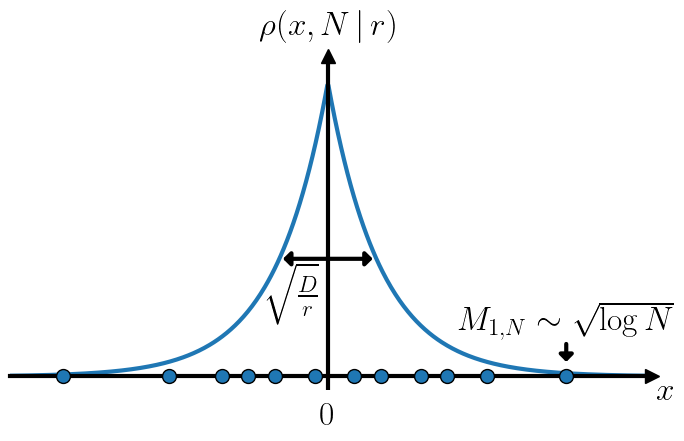}
\caption{The average density $\rho(x,N\vert r)$, given in Eq. \ref{eq:density-A}, is plotted vs $x$ for $N$ simultaneously resetting Brownian motions with $r=D=1$. This density is independent of $N$. The blue dots on the line represent a typical configuration of the particles. The position of the rightmost particle $M_{1,N}$ scales as $\sqrt{\ln N}$ for large $N$.} \label{fig:brownian_example}
\end{figure}

{\bf The NESS and the average density.} In this section, we briefly recall the results of Ref. \cite{Biroli_23} for the simultaneously resetting Brownian gas.  
We consider a gas of $N$ simultaneously resetting Brownian walkers on a line evolving via the stochastic dynamics \ref{reset_dyn} -- see the left panel of Fig. \ref{Fig_intro}. We will see that this is an example of c.i.i.d. variables belonging to the Gumbel class. 
The propagator $p(X \,|\, \tau)$ of a single particle in the absence of resetting, i.e., the probability density to arrive at $X$ at time $\tau$ starting from $X=0$ is simply diffusive, i.e., 
\begin{equation} \label{free-propagator-A}
p(X \,|\, \tau) = \frac{1}{\sqrt{4 \pi D \tau}} \exp[ -\frac{X^2}{4 D \tau} ] \;.
\end{equation}
Plugging this propagator in Eq. \ref{eq:general_NESS} gives the joint distribution in the NESS 
\begin{equation} \label{JPDF-stat}
{\rm Prob.}[\vec{X}]_{\rm NESS} = r \int_0^{+\infty} \frac{\dd \tau}{(4 \pi D \tau)^{N/2}} \exp[ - r \tau - \frac{1}{4 D \tau} \sum_{i = 1}^N X_i^2 ] \;,
\end{equation}
which is manifestly non-factorisable, illustrating the fact that the positions of the particles become correlated in this NESS. 
The origin of these correlations can be traced back to the simultaneous resetting of the particles \cite{Biroli_23}. Given this joint distribution, one can compute various observables in principle. For example, the average density in the steady state is given by Eq. \ref{eq:gen_density} 
\begin{equation} \label{eq:density-A}
\rho(x, N \vert r) = r \int_0^{+\infty} \dd\tau\; e^{-r \tau} p(x \,|\, \tau) = \frac{1}{2} \sqrt{\frac{r}{D}} e^{-\sqrt{\frac{r}{D}} |x|} \;.
\end{equation}
Thus, even though at fixed $\tau$ the marginal distribution of the c.i.i.d. variables is Gaussian, once averaged over the conditioning variable $\tau$,
the steady state average density becomes highly non-Gaussian. A plot of this density is shown in Fig. \ref{fig:brownian_example}.

\vspace{0.5cm}

\noindent{\bf Center of mass.} The two first moments of the distribution $p(X \, |\,  \tau)$ are clearly finite and are given by
\begin{equation} \label{eq:mv_gaussian}
m(\tau) = 0 \mbox{~~and~~} \sigma^2(\tau) = 2 D \tau \;.
\end{equation}
Since $m(\tau) = 0$ we have to use Eq. \ref{CLT-var-res} to obtain the statistics of the center of mass. Plugging Eq. \ref{eq:mv_gaussian} into Eq. \ref{CLT-var-res} we immediately obtain
\begin{equation}
P(C, N) = r \int_0^{+\infty} \dd \tau \; \sqrt{\frac{N}{4 \pi D \tau }} e^{- \frac{N C^2}{4 D \tau}} e^{-r \tau} = \frac{1}{2} \sqrt{\frac{r N}{D}} e^{- \sqrt{\frac{r N}{D}} |C|} \;.
\end{equation}
\begin{figure}[t]
\centering
\begin{minipage}[b]{0.49\textwidth}
\includegraphics[width=\textwidth]{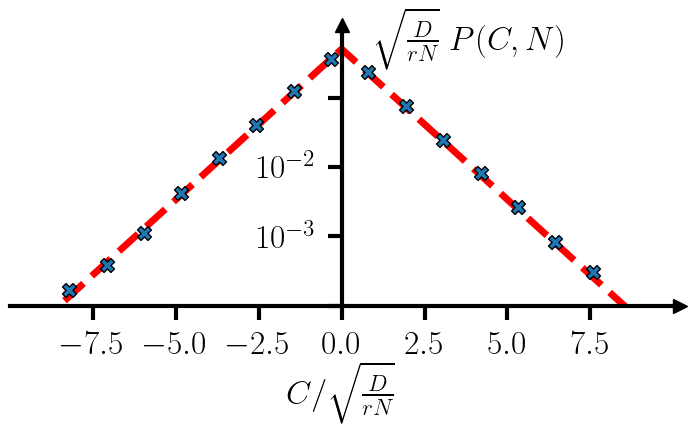}
\end{minipage}
\hfill
\begin{minipage}[b]{0.49\textwidth}
\includegraphics[width=\textwidth]{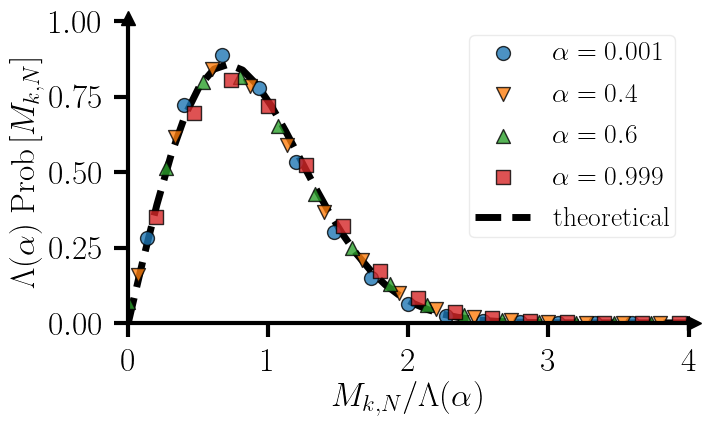}
\end{minipage}
\caption{Plots of the probability density functions of the center of mass (left panel) and the order statistics (right panel) obtained in Eq. \ref{eq:C2rho} and Eq. \ref{eq:res_brownian} respectively. The dashed lines correspond to the theoretical predictions in Eq. \ref{eq:C2rho} -- left panel -- and Eq. \ref{eq:res_brownian} -- right panel. The symbols represent the results of numerical simulations. Different colors and symbols correspond to different values of $\alpha = k/N$, where we used $N=1000$.} \label{scaling_brownian}
\end{figure}
A plot of this scaling function is given in the left panel of Fig. \ref{scaling_brownian} where it is also compared to numerical simulations.
Note that this is simply a rescaling of the average density we derived in Eq. \ref{eq:density-A}.
This is because in this specific case we actually have a nice trick which allows us to express the center of mass exactly for any $N$ in terms of the density function. 
Indeed, making use of the fact that Gaussian variables are stable under summation, i.e., if $X$ and $Y$ are Gaussian then $X + Y$ is also Gaussian, we can write
\begin{equation} \label{eq:psN}
p(S \, |\,  N \tau) = \int \dd X_1 \int \dd X_2 \cdots \int \dd X_N \; p(X_1 \,|\, \tau) p(X_2 \,|\, \tau) \cdots p(X_N \,|\, \tau) \delta(X_1 + X_2 + \cdots + X_N - S) = {\rm Prob.}\left[ \sum_{i = 1}^N X_i = S \right] \;,
\end{equation}
where $S = C\,N$. This formula can be understood as follows. 
In order to reach $S$ in time $N \tau$ we split the trajectory into segments of duration $\tau$. 
Denote by $Z_k$ the position at time $k \tau$ for $0 \leq k \leq N$. 
Then telescopically we can write
\begin{equation}
S = Z_N = (Z_N - Z_{N-1}) + (Z_{N-1} - Z_{N-2}) + \cdots + (Z_1 - 0) \;.
\end{equation}
Then the terms in each parenthesis $X_k = Z_k - Z_{k - 1}$ for $1 \leq k \leq N$ correspond to the distance traveled in the time interval $[(k-1)\tau, k \tau]$. 
Since Gaussian variables are stable under summation we can split the probability on each of these time segments, leading to Eq. \ref{eq:psN}. 
Then we get
\begin{equation}
P(C, N) = r N \int_0^{+\infty} e^{-r \tau} \, {\rm Prob.}\left[\sum_{i = 1}^N X_i = c N  \, \Bigg| \, \tau\right] \dd \tau = r N \int_0^{+\infty} e^{-r \tau} \, p( C N \,|\, N \tau) \; \dd \tau \;.
\end{equation} 
Changing variable to $y = N \tau$ we get
\begin{equation} \label{eq:C2rho}
P(C, N) = N \left[\frac{r}{N} \int_0^{+\infty} e^{-\frac{r}{N} y} \, p( C N \,|\, y) \; \dd y \right] = N \rho\left( C N, N \, \Big| \, r' = \frac{r}{N} \right) \;,
\end{equation}
where $\rho(x, N | r')$ is precisely the average density in Eq. \ref{eq:density-A}.

\vspace{0.5cm}

\noindent{\bf EVS and order statistics.} We now consider the order statistics in the bulk as well as the edge of this gas. The conditional distribution $p(X \vert \tau)$ in Eq. \ref{free-propagator-A} has a Gaussian tail and hence it clearly belongs to the Gumbel class in Eq. \ref{eq:gumbel_tail}. Then, computing the explicit form of the quantile $q(\alpha, \vec{Y}) \equiv q(\alpha, \tau)$ from Eq. \ref{eq:def_wstar}, we obtain
\begin{equation} \label{w-star-sol-A}
q(\alpha, \tau) = \sqrt{4 D \tau} \, {\rm erfc}^{-1}(2 \alpha)\;,
\end{equation}
where ${\rm erf}(z) = 2/\sqrt{\pi} \int_z^{\infty} e^{-u^2}\, \dd u$ and ${\rm erfc}^{-1}(z)$ is the associated inverse function. 
As argued in Section \ref{gumbel}, the order statistics in the large $N$ limit, both in the bulk and at the edges, can be obtained within the same framework, namely from Eq. \ref{eq:res_bulk_integral_form}, with the substitution $\vec{Y} \to \tau$ and $h(\vec{Y}) = r\,e^{-r\tau}$. Plugging Eq. \ref{w-star-sol-A} in \ref{eq:res_bulk_integral_form} gives
\begin{equation} \label{eq:Mk-delta}
{\rm Prob.}[M_{k, N} = w] = r \int_0^{+\infty} \dd\tau e^{-r\tau} \delta\left[w -  \sqrt{4  D \tau} \, {\rm erfc}^{-1}(2\alpha) \right] \;.
\end{equation}
Performing this integral we immediately obtain 
\begin{equation} \label{eq:res_brownian}
{\rm Prob.}[M_{k, N} = w] = \frac{1}{\Lambda(\alpha)} f\left( \frac{w}{\Lambda(\alpha)} \right) \mbox{~~with~~} \Lambda(\alpha) = \sqrt{\frac{4 D}{r}} {\rm erfc}^{-1}(2\alpha) \;,
\end{equation}
where the normalized scaling function $f(z)$ defined on $z > 0$ is given by
\begin{equation}
f(z) = 2 \, z \, e^{-z^2}, \quad z \geq 0 \;,
\end{equation}
This result was already derived in Ref. \cite{Biroli_23}. In the right panel of Fig. \ref{scaling_brownian}, we compare this analytical prediction to numerical simulations, finding excellent agreement.

\subsection{Random ballistic motion: an example of the Weibull class}
\label{ballistic}

\begin{figure}
\centering
\includegraphics[width=0.6\textwidth]{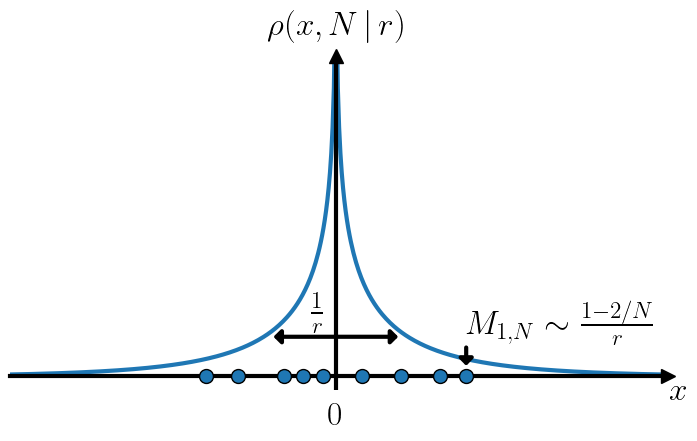}
\caption{The average density $\rho(x,N\vert r)$, given in Eqs. \ref{rho_bal} and \ref{rho_bal1}, is plotted vs $x$ for $N$ simultaneously resetting ballistic particles with $r=1$. This density is independent of $N$. The blue dots on the line represent a typical configuration of the particles. The position of the rightmost particle $M_{1,N}$ is of order $(1-2/N)/r$ for large $N$.} \label{fig:ballistic_example}
\end{figure}

{\bf The NESS and the average density.} To find a physical example belonging to the Weibull class of c.i.i.d. variables, 
we consider a gas of $N$ particles undergoing ballistic motions on the line. The particles all start at the origin and reset simultaneously to the origin
with rate $r$. At the end of every resetting event, each particle is assigned independently a random velocity $v_i$ drawn from a uniform distribution 
$n(v)$  
\begin{equation} \label{eq:def_nv}
n(v) = \begin{dcases}
\frac{1}{2} &\mbox{~~if~~} v \in [-1, 1]\\
0 &\mbox{~~otherwise} \;.
\end{dcases}
\end{equation}
In this case, the free propagator of a single particle at time $\tau$ (in the absence of resetting) is given by 
\begin{equation} \label{free-propagator-C}
p(X \,|\, \tau) = \int_{-\infty}^{+\infty} \dd v \, \delta(X - v \tau) n(v) = \frac{1}{\tau} n\left( \frac{X}{\tau} \right) \;.
\end{equation}
An example of the typical trajectories of this system of particles is shown in the middle panel of Fig. \ref{Fig_intro}. Plugging this free-propagator from Eq. \ref{free-propagator-C} into the general formula for the NESS given in Eq. \ref{eq:general_NESS}, we get
\begin{equation}
{\rm Prob.}[\vec{X}]_{\rm NESS} = r \int_0^{+\infty} \dd\tau \; e^{-r \tau} \frac{1}{\tau^N} \prod_{i = 1}^N n\left( \frac{X_i}{\tau} \right) \;.
\end{equation}
Once again, the particle positions are strongly correlated in the NESS since the joint distribution does not factorize. Given this exact joint distribution, one can compute various observables, as in the Brownian case. For example, the average density is given by  
\begin{equation}\label{rho_bal}
\rho(x, N \vert r) = r \int_0^{+\infty} \dd\tau\; e^{-r \tau} p(x \, |\,  \tau) = \frac{r}{2} \int_{r |x|}^{+\infty} \frac{\dd v}{v} e^{-v} = r \, \rho_s\left( r x \right) \;,
\end{equation}
where the normalized scaling function $\rho_s(z)$ defined for $z \in \mathbb{R}$ is given by
\begin{equation}\label{rho_bal1}
\rho_s(z) = \frac{1}{2} \int_{|z|}^{+\infty} \frac{\dd v}{v} e^{-v} = - \frac{1}{2} {\rm Ei}(-|z|) \;.
\end{equation}
Here, ${\rm Ei}(z)$ is the exponential integral function \cite{Grad} and the scaling function $\rho_s(z)$ is plotted in Fig.  \ref{fig:ballistic_example}. The asymptotics of this scaling function are given by
\begin{equation}
\rho_s(z) \simeq \begin{cases}
- \frac{1}{2} \log z &\mbox{~~for~~} |z| \ll 1\\
e^{-z}/(2 z) &\mbox{~~for~~} |z| \gg 1
\end{cases}\;.
\end{equation}

\vspace{0.5cm}

\begin{figure}
\centering
\begin{minipage}[b]{0.48\textwidth}
\centering
\includegraphics[width=\textwidth]{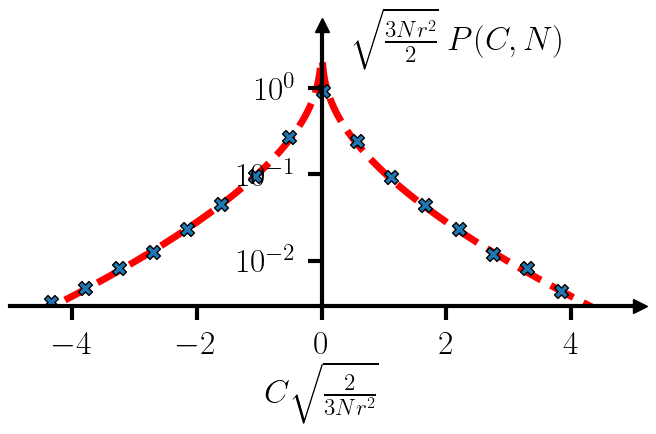}
\end{minipage}\hfill
\begin{minipage}[b]{0.48\textwidth}
\centering
\includegraphics[width=\textwidth]{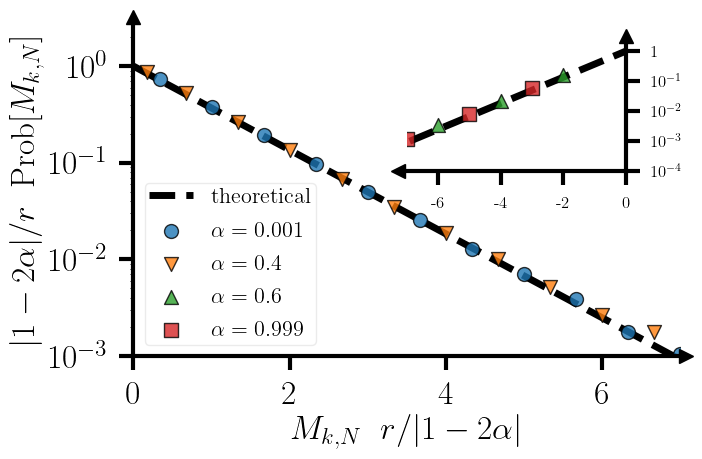}
\end{minipage}
\caption{{\bf Left:} The scaled distribution of the center of mass. The red dotted lines denote the analytical scaling function $\ell(z)$ in Eq. \ref{eq:ell} and the points represent numerical simulations. {\bf Right:} The distribution of the $k$-th maximum $M_{k = \alpha N,N}$ in the bulk, given in Eq. \ref{eq:res_os_ballistic}, is plotted on a log-linear scale. In the main figure, we show the curves for $\alpha = 0.001$ and $\alpha = 0.1$ (where the distribution is supported over the positive semi-axis), while the inset shows the curves for $\alpha = 0.6$ and $\alpha = 0.999$ (where the distribution is supported over the negative semi-axis). The black dotted line represent the analytical prediction given in Eq. \ref{eq:res_os_ballistic} and the symbols represent simulation results for $N=1000$ and $r=1$.} \label{scaling_ballistic}
\end{figure}

\noindent{\bf Center of mass.} The first and second moment in this case are also finite and given by
\begin{equation} \label{eq:mv_ballistic}
m(\tau) = 0 \;,
\end{equation}
and 
\begin{equation}
\sigma^2(\tau) = \int_{-\infty}^{+\infty} \dd X\; X^2 p(X \,|\, \tau) - \left(\int_{-\infty}^{+\infty} \dd X\; X p(X \,|\, \tau)\right)^2 = \int_{-\infty}^{+\infty} \dd X\; \frac{X^2}{\tau} n\left( \frac{X}{\tau} \right) = \int_{-\tau}^{+\tau} \dd X\; \frac{X^2}{2 \tau} = \frac{\tau^2}{3} \;,
\end{equation}
where we used Eqs.\ref{eq:def_nv} and \ref{free-propagator-C} . 
Since $m(\tau) = 0$ is a constant, independent of $\tau$, we can use Eq. \ref{CLT-var-res} to express the p.d.f. of the center of mass. 
Plugging Eq. \ref{eq:mv_ballistic} into Eq. \ref{CLT-var-res} we get
\begin{equation}
P(C, N) = r \int_0^{+\infty} \dd \tau\; \sqrt{\frac{3 N}{2 \pi \tau^2}} \exp[ - \frac{3 N C^2}{2 \tau^2} - r \tau] \;.
\end{equation}
Changing variable to $\nu = r \tau$ we obtain
\begin{equation}
P(C, N) = \sqrt{\frac{3}{2} N r^2} \;\; \ell\left( C \sqrt{\frac{3}{2} N r^2}  \right) \;,
\end{equation}
where the normalized scaling function $\ell(z)$ defined for $z \in \mathbb{R}$ is given by
\begin{equation} \label{eq:ell}
\ell(z) = \frac{1}{\sqrt{\pi}} \int_0^{+\infty} \frac{\dd \nu}{\nu} e^{-\nu - z^2/\nu^2} \;.
\end{equation}
While this integral does not admit a simple closed form, we can easily compute the asymptotics behaviors, which are given by
\begin{equation}
\ell(z) \to \begin{cases}
- \frac{1}{\sqrt{\pi}} \log(z) &\mbox{~~for~~} |z| \ll 1\\
\left( \frac{\sqrt{3} |z|^{1/3}}{2^{1/3}} \right)^{-1} \exp( - \frac{3 |z|^{2/3}}{2^{2/3}} ) &\mbox{~~for~~} |z| \gg 1
\end{cases} \;.
\end{equation}
A plot of this normalized scaling function $\ell(z)$ in Eq. \ref{eq:ell} is given in the left panel of Fig. \ref{scaling_ballistic}, where it is also compared to numerical simulations, finding excellent agreement. 

\vspace{0.5cm}

\noindent{\bf EVS and order statistics.} As in the Gumbel case, one can obtain the distribution of the $k$-th maximum, both in the bulk as well as at the edge, using the same framework leading to Eq. \ref{eq:res_bulk_integral_form}, with the substitution $\vec{Y} \to \tau$ and $h(\vec{Y}) = r\,e^{-r\tau}$. We start by computing the explicit form of the quantile $q(\alpha, \vec{Y}) \equiv q(\alpha, \tau)$. 
Substituting Eq. \ref{free-propagator-C} in Eq. \ref{eq:def_wstar} we obtain
\begin{equation} \label{quantile_weibull}
\alpha = \int_{q(\alpha, \tau)}^{+\infty} p(X \,|\, t) \dd X = \int_{q(\alpha, \tau)}^{+\infty} \frac{\dd X}{\tau} n\left( \frac{X}{\tau} \right) = \int_{q(\alpha, \tau)/\tau}^{+\infty} \dd v \, n(v) \;.
\end{equation}
Using the fact that $n(v)$ in Eq. \ref{eq:def_nv} is supported over the finite interval $v \in [-1,+1]$, we need to consider three cases: if $q(\alpha, \tau) \leq -\tau$ then the integral is equal to $1$, if $q(\alpha, \tau) \geq \tau$ then the integral is equal to 0 and otherwise we have
\begin{equation}
\alpha = \frac{\tau - q(\alpha, \tau)}{2\tau} \; .
\end{equation}
Hence we get
\begin{equation} \label{w-star-sol-C}
q(\alpha, \tau) = \tau(1 - 2 \alpha) \quad {\rm where} \quad 0 \leq \alpha \leq 1 \;.
\end{equation}
The free-propagator, defined in Eq. \ref{free-propagator-C}, clearly belongs to the Weibull class in Eq. \ref{tail_weibull} with $\mu(\vec{Y}) = 1$. 
Therefore, as discussed in Section \ref{Order}, one can once more replace, for large $N$, the  Gaussian distribution in Eq. \ref{eq:Mk_gauss} by a 
delta function, leading to Eq. \ref{eq:res_bulk_integral_form} for the order statistics both in the bulk as well as the edges. 
Plugging Eq. \ref{w-star-sol-C} into Eq. \ref{eq:res_bulk_integral_form} we obtain
\begin{equation} \label{eq:res_os_ballistic}
{\rm Prob.}[M_{k,N} = w] \approx r \int_0^{+\infty} \dd \tau \, e^{- r\tau} \delta\left[w - {\tau} (1 - 2\alpha)\right] = \frac{r}{|1 - 2\alpha|} \exp\left(-\frac{r |w|}{|1 - 2\alpha|} \right) \Theta(w(1 - 2\alpha)) \;.
\end{equation}
Thus the distribution of the $k$-th maximum $M_{k,N}$, both in the bulk as well as at the edges, is given by an exponential distribution, which is   
supported on the positive half-line (resp. negative half-line) for $\alpha < 1/2$ (resp. for $\alpha > 1/2$), as shown in the right panel of Fig. \ref{scaling_ballistic}.

\subsection{L\'evy Flights: an example of the Fr\'echet case}
\label{levy}

\begin{figure}
\centering
\includegraphics[width=0.6\textwidth]{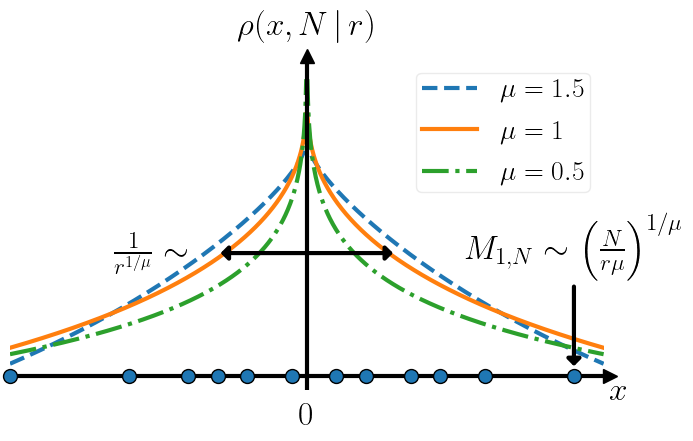}
\caption{The average density $\rho(x,N\vert r)$, given in Eqs. \ref{eq:rho_levy} and \ref{eq:rho_mu}, is plotted vs $x$ for $N$ simultaneously resetting L\'evy flights for three different values of $\mu = 0.5, 1, 1.5$ and $r=1$. This density is independent of $N$. For $\mu=0.5$ and $\mu = 1$, the density diverges as $|z| \to 0$, while it approaches a constant for $\mu  = 1.5$. 
The black dots on the line represent a typical configuration of the particles. The position of the rightmost particle $M_{1,N}$ is of order $N^{1/\mu}$ for large $N$.} \label{fig:levy_example}
\end{figure}

{\bf The NESS and the average density.} In this section, we consider a gas of $N$ L\'evy flights on the line, all starting at the origin and resetting 
simultaneously to the origin. Usually L\'evy flights (in the absence of resetting) are defined in discrete time where the position of a walker evolves via
\begin{equation} \label{xn}
X_{n} = X_{n-1} + \eta_n \quad, \quad X_0 = 0 \;,
\end{equation}
where $\eta_n$'s are i.i.d. random noises with a power-law tail, $p(\eta) \sim |\eta|^{-1 - \mu}$ with $0 < \mu < 2$. For large number of steps $n$, one can replace the discrete time $n$ by a continuous variable $\tau$ and it is well known \cite{Bouchaud-Georges} that the free propagator of the L\'evy flight for large $\tau$ converges to
\begin{equation} \label{free-propagator-B}
p(X \, | \, \tau) \approx \frac{1}{\tau^{1/\mu}} \mathcal{L}_\mu \left( \frac{X}{\tau^{1/\mu}} \right) \;,
\end{equation}
where $\mathcal{L}_\mu(z)$ is the p.d.f. of a centered, normalized to unity, symmetric stable function parametrized by $0<\mu<2$. Usually the function 
$\mathcal{L}_\mu(z)$ is defined by its characteristic function
\begin{equation} \label{eq:def_Lmu}
\hat{\mathcal{L}}_{\mu}(k) = \int_{-\infty}^{+\infty} \dd z\, e^{i k z} \mathcal{L}_\mu(z) = e^{-|k|^\mu} \quad {\rm implying} \quad  \mathcal{L}_\mu(z) = \int_{-\infty}^{+\infty} \frac{\dd k}{2 \pi} \, e^{-i k z - |k|^\mu} \;. 
\end{equation}
A realisation of the trajectories for $N=3$ L\'evy flights generated using the propagator in Eq. \ref{free-propagator-B} 
is shown in Fig.~\ref{Fig_intro} (right panel). Inserting this expression for the free propagator in the general formula for the NESS in Eq. \ref{eq:general_NESS}, we get the joint distribution of the positions of the simultaneously resetting L\'evy flights in the NESS as
\begin{equation}
{\rm Prob.}[\vec{X}]_{\rm NESS} \approx r \int_{0}^{+\infty} \dd \tau\; e^{-r\tau} \prod_{i = 1}^N \frac{1}{\tau^{1/\mu}} \mathcal{L}_\mu\left(\frac{X_i}{\tau^{1/\mu}}\right)\;.
\end{equation}
Given this joint distribution, one can compute various physical observables, as in the two preceding cases. In particular, the averaged density is given by
\begin{equation} \label{eq:rho_levy}
\rho(x, N\vert r) \approx r \int_0^{+\infty} \dd \tau\; e^{-r\tau} \frac{1}{\tau^{1/\mu}}\mathcal{L}_\mu\left( \frac{x}{\tau^{1/\mu}} \right) = r^{1/\mu} \rho_\mu(r^{1/\mu} x) \;,
\end{equation}
where the normalized scaling function $\rho_\mu(z)$ is symmetric and is given by
\begin{equation} \label{eq:rho_mu}
\rho_\mu(z) = \mu |z|^{\mu - 1} \int_0^{+\infty} \dd u \; \frac{1}{u^\mu} e^{-(|z|/u)^\mu} \mathcal{L}_\mu(u) \;.
\end{equation}
One can compute the asymptotic behavior of the average density. For large $|z|$, one finds that, for all $0 < \mu < 2$,  
\begin{equation} \label{rho_large_z}
\rho_\mu(z) \approx \frac{1}{2\pi} \frac{1}{|z|^{1 + \mu}} \quad, \quad |z| \to \infty \;.
\end{equation}
Interestingly, the small $z$ behavior is quite different depending on the value of $\mu$. Indeed, we get, as $|z| \to 0$
\begin{equation} \label{rho_small_z}
\rho_{\mu}(z)  \approx \begin{dcases}
\frac{c_1}{|z|^{1 - \mu}} &\mbox{~~when~~} 0 < \mu < 1 \;,\\
\frac{1}{\pi} (- \log |z|) &\mbox{~~when~~} \mu = 1 \;,\\
\frac{1}{\mu \sin(\pi/\mu)} &\mbox{~~when~~} 1 < \mu < 2 \;,
\end{dcases} 
\end{equation}
where we introduced the constant
\begin{equation}
c_1 = \mu \int_0^{+\infty} \frac{\dd v}{v^{\mu}} \mathcal{L}_\mu(v) \;.
\end{equation}
Using $\mathcal{L}_\mu(v) \to O(1)$ as $v \to 0$ (see Eq. \ref{eq:tail_stable_law_intro}), we see that the constant $c_1$ is well defined 
for $\mu < 1$. Thus the average density diverges as $|z| \to 0$ for $0 < \mu \leq 1$ (but it is still integrable), while it approaches a constant as $|z| \to 0$ for $1 < \mu < 2$.

\vspace{0.5cm}

\noindent{\bf Center of mass.} The stable distribution ${\cal L}_\mu(z)$ has a power-law tail for large $z$ as in Eq. \ref{eq:tail_stable_law_intro}.
\begin{figure}
\centering
\begin{minipage}[b]{0.32\textwidth}
\centering
\includegraphics[width=\textwidth]{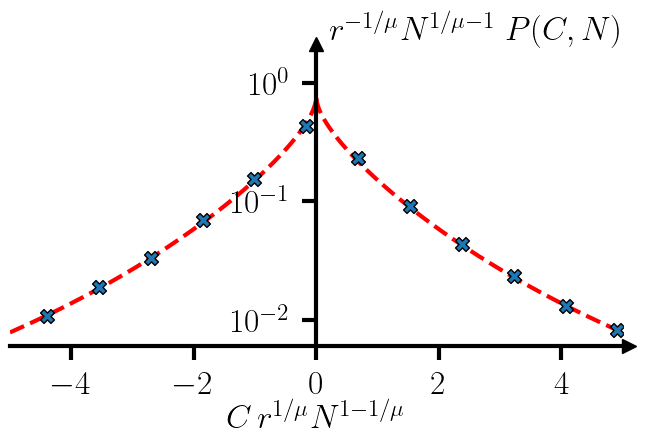}
\end{minipage}
\hfill
\begin{minipage}[b]{0.32\textwidth}
\centering
\includegraphics[width=\textwidth]{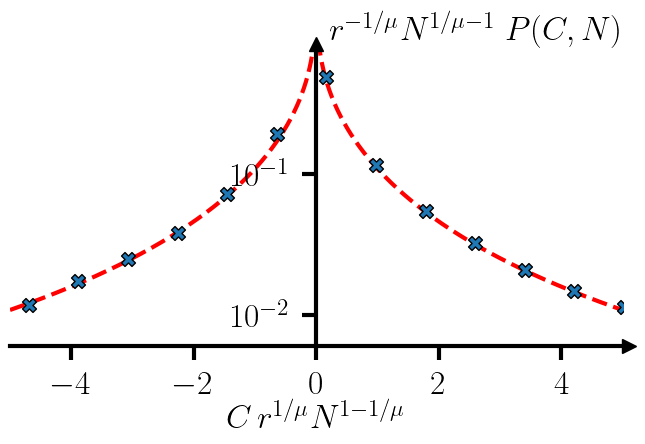}
\end{minipage}
\hfill
\begin{minipage}[b]{0.32\textwidth}
\centering
\includegraphics[width=\textwidth]{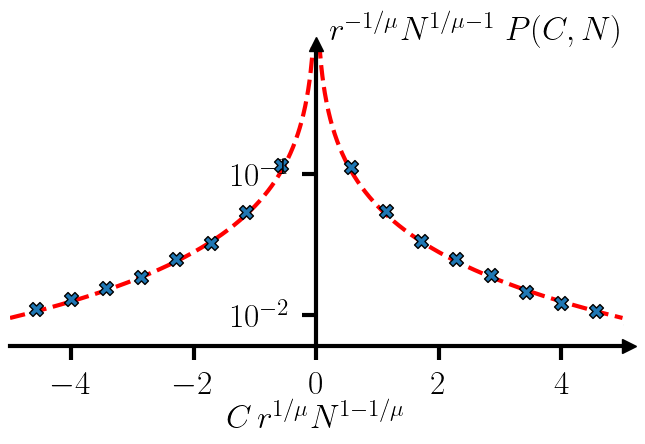}
\end{minipage}
\caption{The p.d.f. of the center of mass $P(C,N)$ in Eq. \ref{eq:res_com_levy} is shown by dashed lines for $N=1000$ and 
for $\mu = 1.5, 1, 0.5$ in the left, middle and right panel respectively. The dots represent numerical simulations.}\label{fig:com_levy}
\end{figure}
Hence, the free-propagator in Eq. \ref{free-propagator-B} belongs to the Fr\'echet class of c.i.i.d. variables. In this case, for the center of mass $C = \frac{1}{N} \sum_{i = 1}^N X_i$, we can use the general result stated in Eq. \ref{CLT-L-res}. A straightforward computation, with suitable change of variables leads us to
\begin{equation} \label{P_C_mu1}
P(C, N) = r N \int_0^{+\infty} \dd \tau \; e^{-r\tau} \frac{1}{(N\,\tau)^{1/\mu}} \mathcal{L}_\mu\left(\frac{C N}{(N\,\tau)^{1/\mu}}\right) \;.
\end{equation}
Making a further change of variable $\nu = N \tau$, one gets  
\begin{equation}\label{P_C_mu2}
P(C, N) = N \left[ \frac{r}{N} \int \dd \nu \; e^{-\frac{r}{N} \nu} \frac{1}{\nu^{1/\mu}} \mathcal{L}_\mu\left(\frac{C N}{\nu^{1/\mu}}\right) \right] = N \rho\left(C N, N \Big| r' = \frac{r}{N} \right) \;,
\end{equation}
where we used the expression for the average density in Eq. \ref{eq:rho_levy}. The fact that this distribution of the center of mass is related to the rescaled single particle propagator can again be traced back to the fact that L\'evy variables are stable under addition. Hence the argument used for the Gaussian case in Eqs. \ref{eq:psN}-\ref{eq:C2rho} can also be applied here, explaining the relation in Eq. \ref{P_C_mu2}. Finally, 
from Eqs. \ref{eq:rho_levy} and \ref{P_C_mu2} we find that the p.d.f. of the center of mass admits the scaling form for large $N$
\begin{equation}
P(C, N) \approx N^{1 - 1/\mu} r^{1/\mu} \rho_\mu( N^{1 - 1/\mu} r^{1/\mu} C) \;, \label{eq:res_com_levy}
\end{equation}
where $\rho_\mu(z)$ is a symmetric function defined in Eq. \ref{eq:rho_mu}. A plot of this scaling function is given in Fig. \ref{fig:com_levy} where it is compared to numerical simulations, showing an excellent agreement.

\vspace{0.5cm}

\noindent {\bf EVS and order statistics.} Unlike in the two cases (Gumbel and Weibull) discussed before, it turns out that for the Fr\'echet case, the order statistics in the bulk can not be extrapolated all the way to the edge. Hence one needs to study separately the statistics 
of $M_{k,N}$ when $k = {\cal O}(N)$ (bulk) and when $k = {\cal O}(1)$ (edge).

\begin{figure}[t]
\centering
\begin{minipage}[b]{0.32\textwidth}
\centering
\includegraphics[width=\textwidth]{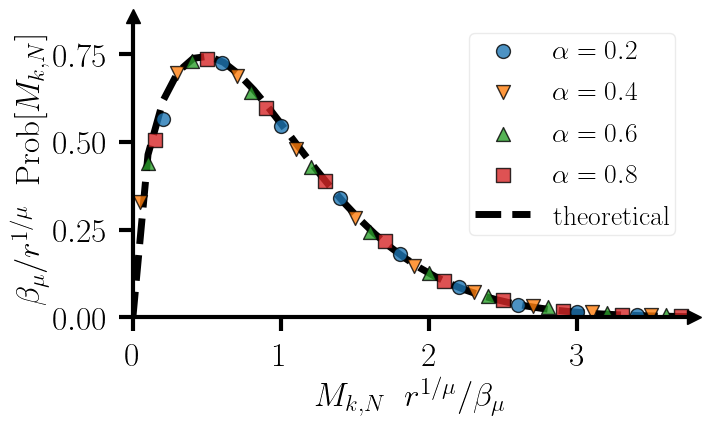}
\end{minipage}
\hfill
\begin{minipage}[b]{0.32\textwidth}
\centering
\includegraphics[width=\textwidth]{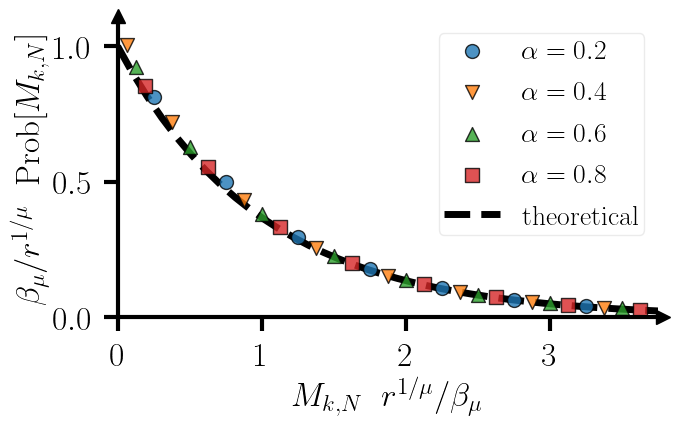}
\end{minipage}
\hfill
\begin{minipage}[b]{0.32\textwidth}
\centering
\includegraphics[width=\textwidth]{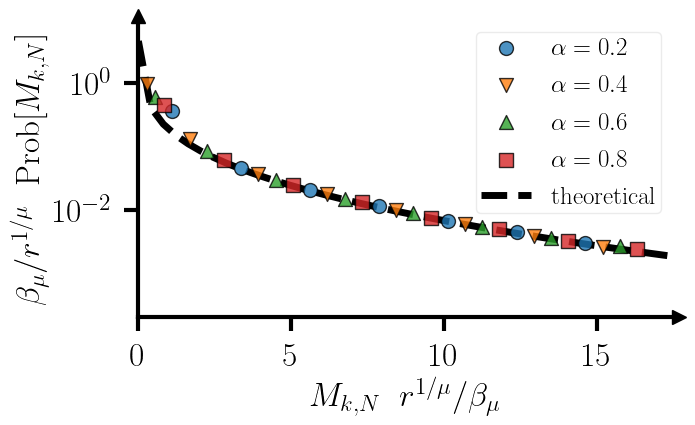}
\end{minipage}
\caption{The p.d.f. of the $k$-th maximum in the bulk $M_{k=\alpha N,N}$ in Eqs. \ref{eq:res_levy_bulk} and \ref{f_mu} is shown for 
$\mu = 1.5, 1, 0.5$ in the left, middle and right panels respectively. Different colors and symbols correspond to different values of $k$. The dashed black line corresponds to the scaling function in Eq. \ref{eq:res_levy_bulk}.}\label{fig:levy}
\end{figure}

\vspace*{0.5cm}
\noindent{\bf Bulk order statistics}. In the bulk, i.e. for $k \sim \mathcal{O}(N)$, we know that the order statistics is given by Eq. \ref{eq:res_bulk_integral_form} with $\vec{Y}$ replaced by $\tau$ and $h(\vec{Y})$ replaced by $r\,e^{-r\tau}$. 
Hence, we start by computing the explicit form of the quantile $q(\alpha, \vec{Y}) \equiv q(\alpha, \tau)$. Replacing Eq. \ref{free-propagator-B} into Eq. \ref{eq:def_wstar} we obtain
\begin{equation} \label{eq_alpha_mu}
\alpha = \int_{q(\alpha, \tau)}^{+\infty} p(X \,|\, \tau)\, \dd X = \int_{q(\alpha, \tau)}^{+\infty} \frac{1}{\tau^{1/\mu}}\mathcal{L}_\mu\left(\frac{X}{\tau^{1/\mu}}\right) \dd X \;.
\end{equation}
Changing variable to $z = X/\tau^{1/\mu}$ and denoting by $F_\mu(z) = \int_{-\infty}^z \mathcal{L}_\mu(x) \dd x$ the cumulative distribution function of the stable law, we can invert the relation in Eq. \ref{eq_alpha_mu} and express it as
\begin{equation} \label{eq:w-star-levy}
q(\alpha, \tau) = \tau^{1/\mu} F_\mu^{-1}(1 - \alpha) = \tau^{1/\mu} \beta_\mu\;,
\end{equation}
where $F_\mu^{-1}$ is the inverse function of $F_\mu$. Although we have no closed form for $\beta_\mu = F_\mu^{-1}(1 - \alpha)$, it is simply a constant which we can therefore numerically compute for any practical purpose. 
Furthermore, from the symmetry of the p.d.f. ${\cal L}_\mu(z)$ it follows that if $\alpha < 1/2$ then $\beta_\mu > 0$, while if $\alpha > 1/2$ then $\beta_\mu < 0$. Exactly at $\alpha = 1/2$, $\beta_\mu = 0$. Hence, in the bulk of the system, in the large $N$ limit, the order statistics will be given by Eq. \ref{eq:res_bulk_integral_form}, which can be simplified as
\begin{equation}\label{eq:res_levy_bulk}
{\rm Prob.}[M_{k, N} = w] \approx r \int_{0}^{+\infty} \dd \tau e^{-r\tau} \delta\left[ w - \tau^{1/\mu} \beta_\mu \right] = \frac{r \mu}{(\beta_\mu)^\mu} w^{\mu - 1} \exp[ - r \left( \frac{w}{\beta_\mu} \right)^{\mu} ] = \frac{r^{1/\mu}}{\beta_\mu} f_\mu \left( r^{1/\mu} \frac{w}{\beta_\mu} \right) \;,
\end{equation}
where the normalized scaling function $f_\mu(z)$ defined for $z > 0$ is given by
\begin{equation}\label{f_mu}
f_\mu(z) = \mu \, z^{\mu - 1} e^{- z^\mu} \,\theta(z) \;,
\end{equation}
and is plotted in Fig. \ref{fig:levy} for three different values of $\mu$. For $\mu > 1$, the scaling function $f_\mu(z)$ vanishes as $z \to 0$, while for $\mu < 1$ it diverges as $z \to 0$.  \\ 

\begin{figure}
\centering
\begin{minipage}[b]{0.32\textwidth}
\centering
\includegraphics[width=\textwidth]{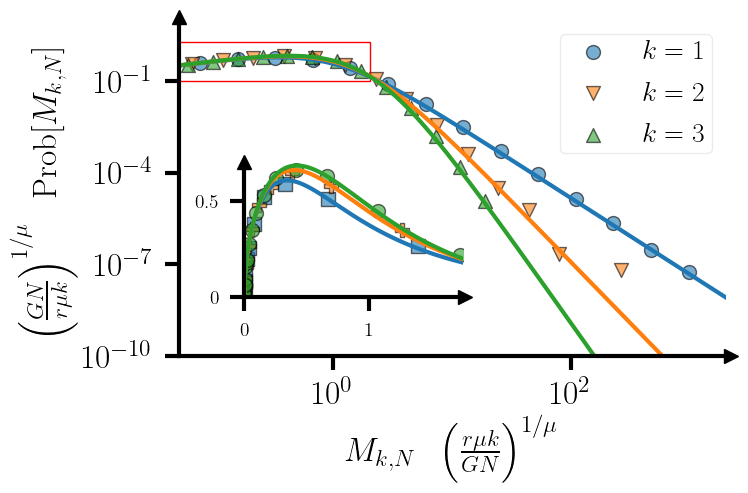}
\end{minipage}
\hfill
\begin{minipage}[b]{0.32\textwidth}
\centering
\includegraphics[width=\textwidth]{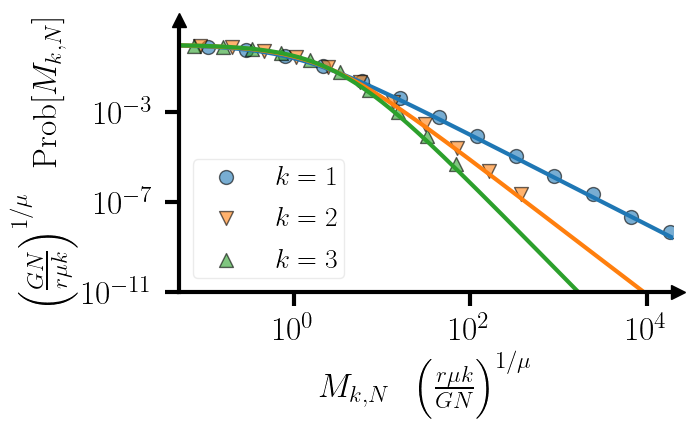}
\end{minipage}
\hfill
\begin{minipage}[b]{0.32\textwidth}
\centering
\includegraphics[width=\textwidth]{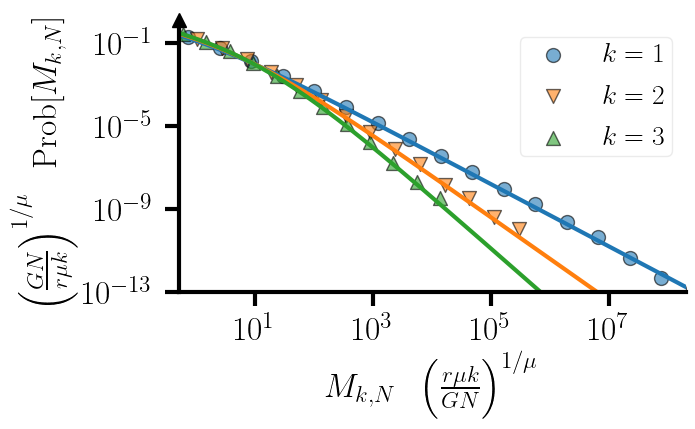}
\end{minipage}
\caption{The p.d.f.'s of the first, second and the third maximum from the right, i.e., $M_{1,N}$, $M_{2,N}$ and $M_{3,N}$, given in Eqs.  \ref{eq:res_levy_edge} and \ref{scaling_Smu} are plotted as solid lines. The three figues correspond to $\mu = 1.5$ (left panel), $\mu = 1$ (middle panel) and $\mu = 0.5$ (right panel). The inset in the left panel shows the behavior close to $z =0$ where it vanishes as $z^{\mu-1} \sim z^{1/2}$. Different colors and symbols correspond to different values of $k$.}\label{fig:levy_edge}
\end{figure}

\vspace*{0.5cm}
\noindent{\bf Edge order statistics}. As argued in the general discussion for the Fr\'echet class in Section \ref{Order}, for the 
order statistics at the edge, we can no longer use the bulk result in Eq. \ref{eq:res_bulk_integral_form}. Instead, we need to use 
Eq. \ref{eq:res:frechet} which holds for any $k = {\cal O}(1)$. To proceed, let us first evaluate the large $X$ behavior of $p(X \, | \, \tau)$
given in Eq. \ref{free-propagator-B}. Using the large $z$ tail of ${\cal L}_\mu(z)$ in Eq. \ref{eq:tail_stable_law_intro} we get 
\begin{equation} \label{large_X_mu}
p(X \,|\, \tau) \underset{X \gg 1}{\sim} \frac{\tau}{X^{1 + \mu}} \sin( \frac{\pi \mu}{2} ) \frac{\Gamma(\mu + 1)}{\pi} \;.
\end{equation}
By replacing $\vec{Y}$ by $\tau$ and $\mu(\vec{Y})$ by $\mu$ in Eq. \ref{eq:tail_frechet}, we see that Eq. \ref{large_X_mu} is a special case of Eq. \ref{eq:tail_frechet}, namely
\begin{equation}\label{large_X_mu_2}
p(X \,|\, \tau) \underset{X \gg 1}{\sim} \frac{G\, \tau}{X^{1 + \mu}}  \;,
\end{equation}
where 
\begin{equation} \label{def_G}
G = \sin( \frac{\pi \mu}{2} ) \frac{\Gamma(\mu + 1)}{\pi} \;.
\end{equation}
Then, using Eq. \ref{large_X_mu_2} and Eq. \ref{eq:def_lambda} it follows that
\begin{equation} \label{lambda_mu}
\lambda_N(w, \tau) = \frac{G \,\tau \,N}{\mu \, w^\mu} \;.
\end{equation}
Hence the random variable $\lambda_N(w, \tau)$ is proportional to the random variable $\tau$, which itself is distributed exponentially via the p.d.f. ${\rm Prob}[\tau] = r \,e^{-r \tau}$. This leads to
\begin{equation}
{\rm Prob.}[\lambda_N(w, \tau) \geq v] = \exp(- r\, \frac{\mu v w^\mu}{G N} ) \;.
\end{equation}
Plugging this expression in Eq. \ref{eq:res:frechet} we then obtain the cumulative distribution function of $M_{k,N}$, which reads
\begin{equation}
{\rm Prob.}[M_{k, N} \leq w] \approx 1 - \frac{1}{\Gamma(k)} \int_0^{+\infty} \dd v \; v^{k-1} \exp( - r\, \frac{\mu v w^\mu}{G N} - v ) = 1 - \frac{1}{[1 + r \mu w^\mu / (G N)]^k} = S_k\left( \frac{r \mu w^\mu}{G N} \right) \;,
\end{equation}
where the scaling function $S_k(z)$ defined for $z \geq 0$ is given by
\begin{equation}
S_k(z) = 1 - \frac{1}{(1 + z)^k} \;.
\end{equation}
The p.d.f. is thus given by
\begin{equation} \label{eq:res_levy_edge}
{\rm Prob.}[M_{k, N} = w] = \frac{r \mu^2 w^{\mu - 1} }{G N} S_k'\left(\frac{r \mu w^\mu}{G N}\right) = \left(\frac{r \mu k}{G N}\right)^{1/\mu} {\cal S}_{\mu,k}\left[  \left( \frac{r \mu k}{G N} \right)^{1/\mu} w  \right] \;,
\end{equation}
where the normalized scaling function ${\cal S}_{\mu,k}(z)$ reads
\begin{equation}\label{scaling_Smu}
{\cal S}_{\mu, k}(z) = \frac{\mu \, z^{\mu - 1}}{(1 + z^\mu/k)^{1 + k}} \quad, \quad z \geq 0 \;.
\end{equation}
For large $z$, the scaling function decays as a power law ${\cal S}_{\mu, k}(z) \sim z^{-1 - \mu\,k}$, while for small $z$, it behaves as ${\cal S}_{\mu, k}(z) \sim z^{\mu -1}$. This scaling function thus characterizes fully the large $N$ behavior of the order statistics at the edge for simultaneously resetting L\'evy flights. In Fig. \ref{fig:levy_edge}, this scaling function is plotted for three different values of $\mu$ in the three panels and compared to numerical simulations, showing a nice agreement. \\

It is instructive to see how this result connects with the order statistics in the bulk. Taking $k = \alpha N$ and performing the limit $N \to +\infty$ keeping $\alpha \sim \mathcal{O}(1)$ fixed yields
\begin{equation}
{\cal S}_{\mu, k}(z) = \mu \, z^{\mu-1} \left[ 1 + \frac{z^\mu}{\alpha N} \right]^{-1 - \alpha N} \underset{N \to +\infty}{\longrightarrow} \mu z^{\mu - 1} e^{-z^\mu} = f_\mu(z)\;.
\end{equation}
Thus, we recover the scaling function obtained for the order statistics in the bulk given in Eq. \ref{eq:res_levy_bulk}. 
The scale factor $r^{1/\mu}/\beta_\mu$ in Eq. \ref{eq:res_levy_bulk} for the bulk also matches the scale factor $\left(\frac{r \mu k}{G N} \right)^{1/\mu}$ in Eq. \ref{eq:res_levy_edge} at the edge. To see this, we first evaluate $\beta_\mu$ in Eq. \ref{eq:w-star-levy}. We get
\begin{equation}
\beta_\mu = F^{-1}_\mu(1 - \alpha) \;,
\end{equation}
which is equivalent to
\begin{equation}
\alpha = \int_{\beta_\mu}^{+\infty} \mathcal{L}_\mu(z) \; \dd z\;.
\end{equation}
In the $\alpha \ll 1$ limit, we can see that $\beta_\mu \gg 1$ and therefore we can use the large $z$ asymptotics of ${\cal L}_{\mu}(z)$ of the stable distribution given in Eq. \ref{eq:tail_stable_law_intro} inside the integral. Then,
\begin{equation}
\alpha \approx \int_{\beta_\mu}^{+\infty} \frac{G}{z^{1 + \mu}} \; \dd z =  \frac{G}{\mu} (\beta_\mu)^{-\mu} \;,
\end{equation}
which yields
\begin{equation}
\beta_\mu \approx \left(\frac{G}{\mu \alpha}\right)^{1/\mu} = \left(\frac{G N}{\mu k}\right)^{1/\mu} \;,
\end{equation}
where we used $\alpha = k/N$. The scale factor in Eq. \ref{eq:res_levy_bulk} is given by
\begin{equation}
\frac{r^{1/\mu}}{\beta_\mu} \underset{\alpha \ll 1}{\longrightarrow} \left(\frac{r \mu k}{G N}\right)^{1/\mu} \;.
\end{equation}
Hence the scale factors in the bulk ands at the edge clearly match in this limit $\alpha \to 0$.

\section{Conclusion}\label{conclusion}

In this paper, we have studied the statistics of the scaled sum and the order statistics 
for a set of $N$ conditionally independent and identically distributed (c.i.i.d.) random variables.
These c.i.i.d. variables are i.i.d. variables for a fixed value of a common parameter (or a set of
parameters), but they get correlated when one averages over the distribution of the parameters.  
This provides a \blue{general} mechanism to generate a large class of strongly correlated random variables that
are still solvable, despite the presence of strong correlations. We studied the analogue of the central limit 
theorem for the scaled sum and also the analogue of the three standard classes of extreme value statistics.  
We showed that averaging over the parameters drastically changes the behavior of these c.i.i.d. observables 
compared to the i.i.d. case. As a physical example, we considered $N$
simultaneously resetting processes on a line where the resetting dynamics makes them strongly correlated, 
but of the c.i.i.d. variety. We showed that when the underlying resetting process is Brownian, this corresponds to
the Gumbel class of c.i.i.d. variables. We introduced a new model of independent ballistic particles with random velocities,
which corresponds to the Weibull class of c.i.i.d. variables. Finally, we considered $N$ simultaneously resetting L\'evy
flights on a line, which corresponds to the Fr\'echet class of c.i.i.d. variables. In all these cases, we obtained exact results
in the large $N$ limit for the distribution of the center of mass (i.e., the scaled sum), as well as for the distribution of the $k$-th maximum (i.e., the order statistics).  

Here, we only considered the statistics of the center of mass and the $k$-th maximum. Using the formalism developed here, one can extend this study to other observables for c.i.i.d. variables, e.g., to the gap between successive particles or the full counting statistics, i.e., the distribution of the number of particles in a given fixed interval. For $N$ simultaneously resetting Brownian motions (Gumbel class), these other observables have been studied recently in \cite{Biroli_23} and it would be interesting to extend these studies to the other two classes, i.e., for the Weibull and the Fr\'echet class. Another interesting direction would be to study these three models in higher dimensions. Furthermore, a special case where the random variables $\vec{X}$ are c.i.i.d. when conditioned on $\vec{Y}$ and the random variables $\vec{Y}$ are themselves c.i.i.d. when conditioned on $\vec{Z}$ might be worth investigating. This hierarchical conditional independence can be iterated and a nice physical example which could be described by this kind of construction is the Generalized Random Energy Model introduced in \cite{GREM}.

\newpage

\blue{\section{Appendix}}

\blue{
In this Appendix, we derive how the central limit theorem of i.i.d. variables generalizes for c.i.i.d. variables. We start by recalling some main results for the sum~\cite{Feller} and the EVS of i.i.d. random variables~\cite{Schehr_14,Majumdar_20}. 
We will generalize these to c.i.i.d. random variables in the following sections. 
Perhaps the most well known result for i.i.d. random variables is the CLT that characterizes the distribution of the scaled sum of $N$ 
i.i.d. random variables in the large $N$ limit. Let us consider a set of $N$ i.i.d. variables $X_1, \cdots, X_N$ each distributed via the probability distribution function (p.d.f.) $p(X)$ that has a finite first and second moment
\begin{equation}
m = \langle X_i \rangle = \int X\, p(X) \dd X 
\end{equation}
and 
\begin{equation}
\sigma^2 = \langle X_i^2 \rangle - \langle X_i \rangle^2 = \int X^2\, p(X) \dd X - \left(\int X\, p(X) \dd X \right)^2 \;,
\end{equation}
where the integrals are over the support of $p(X)$. The CLT states that, for large $N$, the rescaled sum (which we refer to as the ``center of mass'') $C = \frac{1}{N} \sum_{i = 1}^N X_i$, converges to a Gaussian distribution, i.e.,  
\begin{equation} \label{eq:clt}
P(C,N) \underset{N \to +\infty}{\longrightarrow} \sqrt{\frac{N}{2 \pi \sigma^2}} \exp\left( - N \frac{(C - m)^2}{2 \sigma^2} \right) \;,
\end{equation}
centered at $m$ with a variance $\sigma^2/N$. When the p.d.f. $p(X)$ has a diverging second moment, e.g., when $p(X)$ is heavy tailed such as $p(X) \sim X^{-1-\mu}$ for large $X$ with $0< \mu <2$, the centered and scaled sum converges to a L\'evy stable law with index $\mu$~\cite{Feller}.  }

\blue{\subsection{The scaled sum of c.i.i.d. variables}}\label{Sum}

\blue{
In this subsection we calculate the general expressions for the  distribution of the sum of c.i.i.d. variables in the large $N$ limit. 
We find that, in contrast to i.i.d. variables, c.i.i.d. variables in general do not reach universal limit laws akin to the CLT or its generalizations to L\'evy stable laws. We start with the joint distribution in Eq. \ref{def:general2}. The p.d.f. of the sample mean, or equivalently the center of mass, $C=\frac{1}{N} \sum_{i=1}^N X_i$ can be written as
\begin{eqnarray} \label{PC}
P(C,N) = \int \dd {\vec{X}} \; \delta\left(C - \frac{1}{N} \sum_{i=1}^N X_i\right) \; \int \left\{\prod_{i = 1}^N p(X_i | \vec{Y})\right\} h(\vec{Y}) ~\dd \vec{Y} \;.
\end{eqnarray}
Taking the Fourier transform with respect to $C$, one gets
\begin{eqnarray}\label{PC_TF}
\hat P(k,N) = \int P(C,N) \, e^{i k C} \, \dd C = \int \dd \vec{Y} \, \left[\hat p\left( \frac{k}{N}\Big \vert \vec{Y}\right)\right]^N h(\vec{Y}) \;,
\end{eqnarray}
where $\hat{p}(k|\vec{Y})$ is the Fourier transform of the conditional distribution function $p(X|\vec{Y})$ with respect to $X$. }

\blue{
To analyse the large $N$ behavior of $\hat P(k,N)$ in Eq. \ref{PC_TF}, we need to analyse the small $k$ behavior of $\hat p(k\vert \vec{Y})$. 
Let us assume that $p(X|\vec{Y})$ admits a finite first and second moment, 
\begin{equation} \label{moments}
m(\vec{Y}) = \int \dd X \; X \, p(X | \vec{Y}) \mbox{~~and~~} \sigma^2(\vec{Y}) = \int \dd X \; X^2 \, p(X | \vec{Y}) - \left(\int \dd X \; X \, p(X | \vec{Y})\right)^2  \;.
\end{equation}
Then the logarithm of its Fourier transform admits the small $k$ expansion  
\begin{eqnarray}\label{log_p}
\log \hat p(k|\vec{Y}) = i \; m(\vec{Y})\,k - \frac{\sigma^2(\vec{Y})}{2} k^2 + O(k^3) \;. 
\end{eqnarray}
Consequently, exponentiating Eq. \ref{log_p} and keeping terms up to order $O(k^2)$ inside the exponential, one gets
\begin{eqnarray} \label{exp_p}
\hat{p}(k|\vec{Y})\sim e^{i \,m(\vec{Y})\,k -\frac{1}{2}\sigma^2(\vec{Y})k^2} \;.
\end{eqnarray}
Substituting this in Eq. \ref{PC_TF}, and inverting the Fourier transform, we get, for large $N$,
\begin{equation} \label{eq:dnC_theta}
   P(C,N) \approx \frac{1}{2\pi} \int_{-\infty}^{\infty} \dd k \int \dd  \vec{Y}\; e^{i m(\vec{Y}) k -\frac{1}{2 N}\sigma^2(\vec{Y})k^2-ik C}  h(\vec{Y}) \;.
\end{equation}}

\blue{
If $m(\vec{Y})$ is a non-constant function of $\vec{Y}$ then we can drop the quadratic term $O(k^2/N)$ in Eq. \ref{eq:dnC_theta}. Then, to leading order for large $N$, the distribution $P(C,N)$ becomes independent of $N$ with a limiting form  
\begin{equation}
P(C,N) \underset{N \to \infty}{\longrightarrow} \int \dd \vec{Y}\; \delta(m(\vec{Y}) - C) \; h(\vec{Y})\;. \label{CLT_u}
\end{equation}
In particular, the moments of this limiting distribution can be calculated easily from Eq. \ref{CLT_u} leading to
\begin{equation}
    \langle C^n\rangle \underset{N \to \infty}{\longrightarrow} \int \dd \vec{Y} m^n(\vec{Y}) \, h(\vec Y) \;.
\end{equation} 
On the other hand, if $m(\vec{Y}) = m$ is a constant, then one can shift $C$ by $m$ and rescale it by $\sqrt{N}$. It is then easy to see that $P(C,N)$ converges to a scaling form
\begin{eqnarray} \label{scaling_C}
P(C,N) \underset{N \to \infty}{\longrightarrow}  \sqrt{N} \, {\cal P} \left( \left(C-m\right)\sqrt{N}\right)
\end{eqnarray}
where the scaling function ${\cal{P}}(Z)$ is given by
\begin{equation} \label{P_of_Z1}
   {\cal{P}}(Z)  = \frac{1}{2\pi} \int_{-\infty}^{\infty} \dd \tilde k \int \dd \vec{Y} \; e^{-\frac{1}{2}\sigma^2(\vec{Y})\tilde k^2 - i\tilde k Z}  h(\vec{Y}) \;.
\end{equation}
Performing the integral over $\tilde k$, one finds 
\begin{equation} \label{P_of_Z2}
   {\cal{P}}(Z) = \frac{1}{\sqrt{2 \pi}}\int \frac{\dd  \vec{Y}}{\sigma(\vec{Y})} \;\exp \left(-\frac{Z^2}{2\sigma^2(\vec{Y})}\right)\,   h(\vec{Y}) \;.
\end{equation}
This is the main exact result for the limiting distribution of the scaled sum in the case where $m(\vec{Y})$ is independent of $\vec{Y}$. Clearly, the limiting distribution ${\cal{P}}(Z)$ depends on the details of $h(\vec{Y})$ and $\sigma(\vec{Y})$. The moments of the scaling variable $Z$ can be computed from Eq. \ref{P_of_Z2} leading to
\begin{eqnarray}\label{moments_Z}
\langle Z^{2n}\rangle = \frac{\Gamma(2n)}{2^n \Gamma(n)} \int \dd {\vec{Y}} \sigma^{2n}(\vec{Y})\, h(\vec{Y})  \;, \quad {\rm and} \qquad  \langle Z^{2n+1}\rangle=0 \qquad n=0,1,2 \ldots \;.
\end{eqnarray}
Using this result together with the scaling form in \ref{scaling_C} we get
\begin{eqnarray}\label{moments_C}
\langle (C-m)^{2n}\rangle = \frac{\Gamma(2n)}{(2N)^n \Gamma(n)} \int \dd {\vec{Y}} \sigma^{2n}(\vec{Y})\, h(\vec{Y})  \;, \quad {\rm and} \qquad  \langle (C-m)^{2n+1}\rangle=0 \qquad n=0,1,2 \ldots \;.
\end{eqnarray}
}

\blue{
So far, we have assumed that the two first moments of $p(X\vert \vec{Y})$ are finite. In case they are divergent, one can perform a similar analysis 
for c.i.i.d. variables as in the case of i.i.d. L\'evy variables. For simplicity, we assume that the variable $X$ is symmetric with zero mean and the conditional p.d.f. $p(X|\vec{Y})$ has a power law tail $p(X|\vec{Y}) \sim 1/X^{1+\mu}$ for large $X$, with $0 < \mu < 2$. We also assume for simplicity that $\mu$ is independent of $Y$. In this case, one can approximate the small $k$ behavior of the Fourier transform $\hat p(k\vert \vec{Y}) \approx e^{- \,|b(\vec{Y})\, k|^{\mu}}$ where $b(\vec{Y})$ is a scale factor. Substituting this in Eq.~\ref{PC_TF} and inverting the Fourier transform, we get
\begin{eqnarray}\label{PC_TF_mu}
P(C,N) \approx \int_{-\infty}^\infty \frac{\dd k}{2 \pi}  \int \dd \vec{Y} e^{- i k C} e^{-  N^{1-1/\mu} |b(\vec{Y})\, k|^\mu} h(\vec{Y}) \;.
\end{eqnarray}
Performing the change of variable $k = \tilde k N^{1-1/\mu}$, one finds that $P(C,N)$ takes the scaling form
\begin{eqnarray}\label{scaling_mu} 
P(C,N) \approx N^{1-1/\mu} \; \tilde{\cal P}_\mu\left(\frac{C}{N^{1/\mu - 1}}\right)
\end{eqnarray}
where the scaling function $\tilde{\cal P}_\mu(Z)$ reads
\begin{eqnarray}\label{scaling_mu2}
\tilde{\cal P}_\mu(Z) = \int_{-\infty}^\infty \frac{\dd \tilde k}{2 \pi} \int  \dd \vec{Y} e^{- i \tilde k Z} e^{- |b(\vec{Y})\,\tilde k|^\mu} h(\vec{Y}) \;.
\end{eqnarray}
Finally, performing the integral over $\tilde k$, it can be expressed in the compact form
\begin{eqnarray}\label{scaling_mu3}
\tilde{\cal P}_\mu(Z) = \int \frac{\dd \vec{Y}}{b(\vec{Y})} {\cal L}_{\mu} \left( \frac{Z}{b(\vec{Y})}\right) \, h(\vec{Y}) \;,
\end{eqnarray}
where ${\cal L}_{\mu}(z)$ is the L\'evy stable distribution (scaled to unity) as stated in Eq. \ref{stable}. In the case $\mu=2$, ${\cal L}_2(z) = e^{-z^2/4}/(2 \sqrt{\pi})$ and Eq. \ref{scaling_mu3} gives back \ref{P_of_Z2} with $\sigma(\vec{Y}) = \sqrt{2}\,b(\vec{Y})$. 
}
\end{document}